\documentclass[11pt]{article}
\usepackage{jcapmod}

\usepackage{booktabs}
\usepackage[english]{babel}
\usepackage{amsmath,amssymb,amsbsy,amstext, amsthm, simplewick}
\usepackage{hyperref}
\usepackage{graphicx}
\usepackage{amsfonts}
\usepackage{amssymb}
\usepackage[small]{caption}
\usepackage{upgreek}
  \usepackage{framed}
\usepackage{enumitem}
  \usepackage{fontawesome}
  \usepackage{jcapmod}
\usepackage{booktabs}
\usepackage[english]{babel}
\usepackage{amsmath,amssymb,amsbsy,amstext, amsthm, simplewick}
\usepackage{wrapfig}
\usepackage{hyperref}
\usepackage{graphicx}
\usepackage{amsfonts}
\usepackage{amssymb}
\usepackage{subfig}
\usepackage{upgreek}
 \usepackage{exscale,relsize}
 \usepackage[makeroom]{cancel}
\usepackage{soul}
\usepackage{bbold}
\usepackage{lipsum}
\usepackage{mdframed}
\usepackage{mathtools}
\usepackage[dvipsnames]{xcolor}
\usepackage{tikz}
\usepackage{tikz-cd}
\usepackage[export]{adjustbox}

 \newcommand{\circled}[2][]{%
  \tikz[baseline=(char.base)]{%
    \node[shape = circle, draw, inner sep = 1pt]
    (char) {\phantom{\ifblank{#1}{#2}{#1}}};%
    \node at (char.center) {\makebox[0pt][c]{#2}};}}
\robustify{\circled}

\newcommand*\diff{\mathop{}\!\mathrm{d}}
\newcommand{\overbar}[1]{\mkern 1.5mu\overline{\mkern-1.5mu#1\mkern-1.5mu}\mkern 1.5mu}

\usepackage{colortbl}
\definecolor{lightgreen}{cmyk}{0.2, 0, 0.2, 0.2}
\definecolor{lightgray}{cmyk}{0.1,0.2,0,0.1}
\definecolor{lightgray2}{cmyk}{0.1,0.1,0,0.1}

\makeatletter
\newlength{\apb@width}
\newcommand{\autoparbox}[2][c]{\settowidth{\apb@width}{#2}\parbox[#1]{\apb@width}{#2}}

\newcommand{\Cen}[2]{%
  \ifmeasuring@
    #2%
  \else
    \makebox[\ifcase\expandafter #1\maxcolumn@widths\fi]{$\displaystyle#2$}%
  \fi
}
\makeatother

\newcommand{\beq}{\begin{equation}\begin{aligned}}
\newcommand{\eeq}{\end{aligned}\end{equation}}

\newcommand{\CL}{\texttt{$\mathcal{C}\text{osmo}\mathcal{L}\text{attice}$}}

\numberwithin{equation}{section}

\def\beq{\begin{equation}}
\def\eeq{\end{equation}}

\def\Beq{\begin{equation}\begin{aligned}}
\def\Eeq{\end{aligned}\end{equation}}

\def\bea{\begin{eqnarray}}
\def\eea{\end{eqnarray}}

\def\beq{\begin{equation}}
\def\eeq{\end{equation}}
\def\bea{\begin{eqnarray}}
\def\eea{\end{eqnarray}}

\def\bp{\boldsymbol{p}}

\def\bk{\boldsymbol{k}}

\def\bx{\boldsymbol{x}}

\DeclareRobustCommand{\SkipTocEntry}[4]{}
\DeclareSymbolFont{extraup}{U}{zavm}{m}{n}
\DeclareMathSymbol{\varheart}{\mathalpha}{extraup}{86}
\DeclareMathSymbol{\vardiamond}{\mathalpha}{extraup}{87}

\begin{document}

\hypersetup{pageanchor=false}

\begin{titlepage}

\setcounter{page}{1} \baselineskip=15.5pt \thispagestyle{empty}

\bigskip\

\begin{flushright}
\mbox{}
DESY-23-075 
\end{flushright}

\vspace{0.3cm}
\begin{center}

{\fontsize{20.74}{24}\selectfont  \sffamily \bfseries Reheating after Inflaton Fragmentation}

\end{center}

\vspace{0.2cm}

\begin{center}
{\fontsize{12}{30}\selectfont  Marcos A.~G.~Garcia$^{\spadesuit}$\footnote{marcos.garcia@fisica.unam.mx} and Mathias Pierre$^{\clubsuit}$\footnote{mathias.pierre@desy.de}}
\setcounter{footnote}{0}
\end{center}

\begin{center}

\vskip 7pt

\textsl{$^{\spadesuit}$ Departamento de F\'isica Te\'orica, Instituto de F\'isica, Universidad Nacional Aut\'onoma de M\'exico, Ciudad de M\'exico C.P. 04510, Mexico}\\
\textsl{$^{\clubsuit}$ Deutsches Elektronen-Synchrotron DESY, Notkestr. 85, 22607 Hamburg, Germany}
\vskip 7pt

\end{center}

\vspace{0.3cm}
\centerline{\bf ABSTRACT}

\vspace{0.3cm}
\noindent In the presence of self-interactions, the post-inflationary evolution of the inflaton field is driven into the non-linear regime by the resonant growth of its fluctuations. The once spatially homogeneous coherent inflaton is converted into a collection of inflaton particles with non-vanishing momentum. Fragmentation significantly alters the energy transfer rate to the inflaton's offspring during the reheating epoch. In this work we introduce a formalism to quantify the effect of fragmentation on particle production rates, and determine the evolution of the inflaton and radiation energy densities, including the corresponding reheating temperatures. For an inflaton potential with a quartic minimum, we find that the efficiency of reheating is drastically diminished after backreaction, yet it can lead to temperatures above the big bang nucleosynthesis limit for sufficiently large couplings. In addition, we use a lattice simulation to estimate the spectrum of induced gravitational waves, sourced by the scalar inhomogeneities, and discuss detectability prospects. We find that a Boltzmann approach allows to accurately predict some of the main features of this spectrum.
\vspace{0.2in}

\begin{flushleft}
{June} 2023
\end{flushleft}

 \end{titlepage}

\hypersetup{pageanchor=true}

\tableofcontents

\section{Introduction}

Ever since the inception of cosmic inflation as a potential solution to the initial condition problems of standard Big Bang cosmology, the question of how the universe transitions from a cold, empty, quasi-de Sitter state to a radiation dominated stage in thermal equilibrium, has been an active research topic.  The original inflationary proposal ({\em old inflation}) was in particular abandoned due to its impossibility to lead to a phenomenologically successful {\em reheating} of the universe~\cite{Guth:1980zm,Sato:1980yn,Guth:1982pn,Olive:1989nu}. The overwhelming majority of the subsequent proposals, based on the slow-roll of an elementary scalar field called the inflaton, incorporate reheating mechanisms tied to the coherent oscillation of this inflaton field about its minimum.\footnote{The assumption of a dominant classical, coherent component to the inflaton is also present in oscillationless reheating mechanisms such as instant preheating~\cite{Felder:1998vq,Felder:1999pv}.} 

At the perturbative (Boltzmann) level, reheating is modeled as the dissipation of the oscillating inflaton into elementary particles, with a decay rate determined by the quantum mechanical transition amplitude from the time-dependent vacuum to the corresponding particle states. This amplitude is evaluated by averaging the transition rate over the fast oscillation of the inflaton about its minimum, assuming therefore that particle production evolves adiabatically following the expansion of the universe. For two-body decays, the dissipation rate is mostly sensitive to the time-dependence of the effective mass of the oscillating inflaton, which can lead to time-dependent kinematic blocking effects, and in general to a decreasing (or increasing) decay efficiency depending on the shape of the minimum of the inflaton potential~\cite{Shtanov:1994ce,Ichikawa:2008ne,Nurmi:2015ema,Kainulainen:2016vzv,Garcia:2020eof,Garcia:2020wiy}. 

It is known, however, that the perturbative picture is insufficient to describe the dynamics of reheating in the presence of strong couplings, or when the short time-scale oscillation of the inflaton field leads to the resonant enhancement (bosons) or suppression (fermions) of the growth of the quantum fields associated with the decay products. When active, these collective effects, known as {\em preheating}, generically lead to a significant modification of the reheating dynamics~\cite{Dolgov:1989us,Traschen:1990sw,Kofman:1994rk,Shtanov:1994ce,Boyanovsky:1995ud,Yoshimura:1995gc,Kofman:1997yn}. In scenarios with a quadratic inflaton potential, the parametric resonance of the momentum modes of the produced fields can lead to the exponential growth of perturbations in the early stages of reheating. This growth is transient and disordered, due to a mismatch between the particle production rates and the expansion rate of the universe. Nevertheless, if the resonance is sufficiently strong and maintained for a sufficient amount of time, it can bring the growth of fluctuations into the non-linear regime. The inflaton condensate is fragmented in favor of a quasi-thermal bath of inflaton particles and its decay products~\cite{Garcia-Bellido:2002fsq,Felder:2006cc,Frolov:2010sz,Amin:2014eta,Garcia:2021iag}. In some scenarios, the result is that the Boltzmann approximation correctly describes the production of daughter fields only in a very narrow range of the available parameter space~\cite{Garcia:2022vwm}.

Preheating in non-quadratic minima has also been studied extensively~\cite{Greene:1997fu,Kaiser:1997mp,Garcia-Bellido:2008ycs,Frolov:2010sz,Amin:2011hj,Hertzberg:2014iza,Figueroa:2016wxr,Lozanov:2016hid,Lozanov:2017hjm,Fu:2017ero,Antusch:2021aiw,Lebedev:2022vwf,Cosme_2023,Alam:2023kia}. Notably, the presence of inflaton self-interactions, even if small, can accumulate over several oscillations and source the resonant growth of non-zero momenta inflaton modes, eventually backreacting with the homogeneous component. This {\em self-fragmentation} of the inflaton not only re-distributes the initial energy density of the field, but it can also lead to the formation of localized, soliton-like objects such as oscillons~\cite{Copeland:1995fq,Salmi:2012ta,Amin:2010dc,Amin:2011hj,Lozanov:2016hid,Lozanov:2017hjm}. The large inhomogeneities triggered by the self-fragmentation of the inflaton can also efficiently source gravitational wave (GW) production. The typical frequency for the resulting stochastic gravitational wave background is larger than its counterpart sourced during inflation $f\gtrsim\,\text{MHz}$~\cite{Dufaux:2007pt,Garcia-Bellido:2007nns,Ringwald:2022xif}. Such high frequency is beyond the reach of current future earth or space based interferometers such as LISA or the DECIGO but might be in the reach of future dedicated experiments~\cite{Herman:2022fau}. The resulting gravitational wave spectrum would carry precious information about the dynamics of the inflaton in the first instants following inflation and shed light on the reheating epoch.\footnote{Consequences of the presence of a scalar spectator field on the GW spectrum has been explored in several works such as Refs.~\cite{Figueroa:2017vfa,Fu:2017ero}}

The depletion of the coherent condensate eventually shuts down the parametric resonance. Following fragmentation, one must therefore follow the population of the relativistic bath of decay products by perturbative means, accounting for both the inflaton zero and non-zero mode dissipation. The decay rates will differ from those prior to the backreaction epoch, significantly modifying any estimates for the reheating temperature of the universe. A formalism to estimate the evolution of the radiation energy density during reheating in the pre- and post-fragmentation regimes, including the reheating temperature is the main goal of this work. To showcase the differences with respect to the purely perturbative evolution of the inflaton-radiation sector we specialize to the case of a quartic inflaton potential near its minimum.

This paper is organized as follows. In Section~\ref{sec:quartic} we discuss in detail the evolution of an inflaton field in a quartic potential during reheating, in the absence of interactions with other fields. Sec.~\ref{sec:coherent} is devoted to the dynamics of the background evolution of the homogeneous inflaton condensate. In Sec.~\ref{sec:parametricresonance} we study the parametric growth of inflaton fluctuations at linear order, and in Sec.~\ref{sec:backreaction} we study the non-linear regime of their growth, including backreaction and fragmentation effects. For Sec.~\ref{sec:boltzmann} we include the perturbative Boltzmann determination of the spectrum of inflaton fluctuations. The inflaton-matter/radiation couplings are introduced in Section~\ref{sec:reheating}. The analysis is divided into the study of the dissipation of the coherent inflaton condensate (Sec.~\ref{sec:coherentdecay}) and the decay of the inflaton particles after fragmentation (Sec.~\ref{sec:fragdecay}). Our main results, the corresponding reheating temperatures, are presented in Sec.~\ref{eq:rehtemp}. Finally, in Sec.~\ref{sec:gws} we determine the induced gravitational waves from the non-linear scalar dynamics, and discuss their potential for detectability. Our conclusions are presented in Sec.~\ref{sec:conclusions}.

\section{Post-inflationary dynamics in a quartic potential}\label{sec:quartic}

The current measurements of the primordial curvature power spectrum, and bounds on the tensor power spectrum, are compatible with the presence of a single, slowly rolling neutral scalar field which drives inflation. We denote this inflaton by $\phi$, and for it we will assume the following form for the action,
\beq
\label{eq:actionphi}
\mathcal{S} \;=\; \int \diff ^4x\,\sqrt{-g} \left[\frac{1}{2}(\partial_{\mu}\phi)^2 - V(\phi) + \mathcal{L}_{\rm int}\right] \, .
\eeq
Here  $g \equiv \det(g_{\mu \nu})$ is the metric determinant of a flat Friedmann-Robertson-Walker metric with scale factor $a$, $V(\phi)$ denotes the inflaton potential, and $\mathcal{L}_{\rm int}$ denotes its interactions with the rest of the (extended) Standard Model. The potential $V(\phi)$ must be chosen in order to match the measured amplitude and tilt of the scalar power spectrum, $A_{S*}\simeq 2.1\times 10^{-9}$, $n_s\simeq 0.966$, and to avoid the upper bound on the tensor-to-scalar ratio, $r<0.036$~\cite{Planck:2018vyg,Planck:2018jri,BICEP:2021xfz}. Moreover, in this work we impose the condition that this potential is quartic during the post-inflationary reheating. Among several suitable candidates, we choose for definiteness the quartic T-model attractor~\cite{Kallosh:2013hoa,Garcia:2020wiy},
\begin{align}
\label{eq:phipotential}
V(\phi) \;&=\; \lambda M_P^{4} \left[ \sqrt{6}\tanh\left(\frac{\phi}{\sqrt{6}M_P}\right)\right]^4 \\
&\simeq\; \lambda \phi^4\,, \qquad (\phi\ll M_P)
\end{align}
where $M_P = 1/\sqrt{8\pi G_N}\simeq 2.435\times 10^{18}\,{\rm GeV}$ denotes the reduced Planck mass. The constant $\lambda$ is determined by the scalar spectrum amplitude, evaluated at the horizon exit time of the {\em Planck} pivot scale, $k_*=0.05\,{\rm Mpc}^{-1}$. For a generic potential, the number of $e$-folds before the end of inflation at horizon exit time are determined from the expression~\cite{Liddle:2003as,Martin:2010kz}
\begin{align}\label{eq:Nstar} \notag
N_* \;=\; &\ln\left[\frac{1}{\sqrt{3}}\left(\frac{\pi^2}{30}\right)^{1/4}\left(\frac{43}{11}\right)^{1/3}\frac{T_0}{H_0}\right]-\ln\left(\frac{k_*}{a_0H_0}\right) - \frac{1}{12}\ln g_{\rm reh}\\
&\qquad  + \frac{1}{4}\ln\left(\frac{V_*^2}{M_P^4\rho_{\rm end}}\right) + \frac{1-3w_{\rm int}}{12(1+w_{\rm int})}\ln\left(\frac{\rho_{\rm rad}}{\rho_{\rm end}}\right)\,.
\end{align}
Here $H_0=67.36\,{\rm km}\, {\rm s}^{-1}{\rm Mpc}^{-1}$~\cite{Planck:2018vyg}, $T_0=2.7255\,{\rm K}$~\cite{Fixsen:2009ug} and $a_0=1$ denote the present Hubble parameter, photon temperature and scale factor, respectively. The energy density at the end of inflation is denoted by $\rho_{\rm end}$, and the energy density at the beginning of the radiation dominated era by $\rho_{\rm rad}$. The effective number of degrees of freedom during reheating is denoted by $g_{\rm reh}$. The $e$-fold averaged equation of state parameter during reheating corresponds to 
\begin{equation}
w_{\rm int} \;\equiv\; \frac{1}{N_{\rm rad}-N_{\rm end}}\int_{N_{\rm end}}^{N_{\rm rad}} w(n)\,\diff n\,.
\end{equation}
As is well know, and we show explicitly in the following section, for quartic reheating $w\simeq 1/3$. The last term in (\ref{eq:Nstar}) vanishes, and therefore the number of $e$-folds is uniquely determined, $N_*\simeq 56$ for T-model inflation, and so is the coupling constant $\lambda\simeq3.3\times 10^{-12}$. We now study the post-inflationary evolution of the inflaton sector, from the coherent oscillation stage, to the backreaction and post-fragmentation phase.

\subsection{Coherent oscillations}\label{sec:coherent}

We first consider the dynamics of the inflaton field at the background level after the end of inflation. Inflation ends when $\ddot{a}=0$, or equivalently when $\dot{\phi}^2=V(\phi)$, which for the T-model (\ref{eq:phipotential}) corresponds to 
\beq
\phi_{\rm end} \;\simeq\;1.52\,M_P \,, \quad \rho_{\rm end}\;\simeq\; (4.5\times 10^{15}\,{\rm GeV})^4\,.
\eeq
Our focus here will be the early stage of reheating, with $a/a_{\rm end}\lesssim \mathcal{O}(10^2)$. Importantly, we will assume that the main decay channel of the inflaton is to fermionic fields, which do not manifest the exponential growth from the parametric resonance of their mode equations. This allows us to assume that the coherence of the inflaton oscillation is maintained, and that the third term in (\ref{eq:actionphi}) can be disregarded at early times~\cite{Garcia:2020eof,Garcia:2020wiy}. If this is the case, variation of the action with respect to the homogeneous inflaton field and metric yields the equations of motion
\begin{align}\label{eq:eomback}
\ddot{\phi} + 3H\dot{\phi} + 4\lambda\phi^3 \;&\simeq\;0\,,\\
\frac{1}{2}\dot{\phi}^2 + \lambda\phi^4\;&\simeq\; 3M_P^2H^2\,,
\end{align}
where $H=\dot{a}/a$ is the Hubble parameter, and a dot represents differentiation with respect to cosmic time. These equations describe an underdamped anharmonic oscillator, which can be parametrized in terms of an envelope function $\phi_0(t)$, encoding the redshift due to expansion, and a quasi-periodic function $\mathcal{P}(t)$, which encodes the short time-scale oscillation, 
\beq\label{eq:pdef}
\phi(t) \;\simeq\; \phi_0(t)\,\mathcal{P}(t)\,.
\eeq
 To determine the time-dependence of the envelope function, we multiply Eq.~(\ref{eq:eomback}) by $\phi$ and average over one oscillation. This yields,
\beq
\langle \phi\ddot{\phi}  + 3 H \phi \dot{\phi}  + 4\lambda \phi^4 \rangle \;\simeq\; -\langle \dot{\phi}^2\rangle + 4\langle V(\phi)\rangle \;=\; 0\,,
\label{eq:averageEOMinflaton}
\eeq
implying that the oscillation-averaged energy and pressure densities can be written as~\cite{Garcia:2020wiy}
\begin{align}\label{eq:rhoback}
\rho_{\phi} \;&\equiv\; \frac{1}{2}\langle \dot{\phi}^2\rangle + \langle V(\phi)\rangle \;\simeq\; 3\langle V(\phi)\rangle \;=\; V(\phi_0)\,,\\
p_{\phi} \;&\equiv\; \frac{1}{2}\langle \dot{\phi}^2\rangle - \langle V(\phi)\rangle \;\simeq\; \langle V(\phi)\rangle \;=\; \frac{1}{3} V(\phi_0)\,,
\end{align}
where we have used $\langle \mathcal{P}^4\rangle = \frac{1}{3}$, as it can be checked from the solution for $\mathcal{P}$, see (\ref{eq:eqforP}) below.  Hence, the equation of state parameter is $w_{\phi} \simeq 1/3$, and the equation of motion (\ref{eq:eomback}) can be rewritten as 
\beq
\dot{\rho}_{\phi} + 4H\rho_{\phi} \;\simeq\;0\,.
\eeq
Thus, as is well known, the energy density of the inflaton redshifts as radiation, $\rho_{\phi} \propto a^{-4}$. As a result, we determine that the decaying envelope of the oscillation redshifts as
\beq\label{eq:phiredshift}
\phi_0(t) \;\simeq\; \phi_{\rm end} \left(\frac{a_{\rm end}}{a(t)}\right)\,.
\eeq
The time-dependence of the quasi-periodic function $\mathcal{P}(t)$ can be obtained by exploiting the underdamped nature of the oscillations. Approximating the envelope as constant during one oscillation, the equation of motion (\ref{eq:eomback}) can be written as
\beq\label{eq:eqforP}
\dot{\mathcal{P}}^2 \;=\; \frac{1}{6}m_{\phi}^2\left(1-\mathcal{P}^4\right)\,.
\eeq
where the effective mass of the inflaton is defined as
\beq\label{eq:mphi}
m_{\phi}^2(t) \;\equiv\; V_{\phi \phi} (\phi_0(t)) \;=\; 12\lambda \phi_0^2(t) \;\simeq\; 12\lambda \phi_{\rm end}^2 \left(\frac{a_{\rm end}}{a(t)}\right)^2 \;\equiv\; m_{\rm end}^2 \left(\frac{a_{\rm end}}{a(t)}\right)^2\,,
\eeq
with $V_{\phi \phi}=\partial^2 V/\partial \phi^2$. Equation (\ref{eq:eqforP}) can be solved in terms of Jacobi elliptic functions, $\mathcal{P}(t) \;=\; {\rm sn}\, ( m_{\phi}t/\sqrt{6},-1)$ which can be expanded in terms of Fourier coefficients as
\begin{equation}
\mathcal{P}(t)=\sum_{n=-\infty}^{\infty} \mathcal{P}_{n} e^{-i n \omega_\phi t} \,,
\label{eq:expansionP}
\end{equation}
where the sum is taken over the harmonic modes of the $\phi$ oscillation with the fundamental angular frequency given by
\beq\label{eq:omegaP}
\omega_{\phi} \;=\; m_{\phi}\sqrt{\frac{2}{3}\pi}\, \frac{\Gamma(3/4)}{\Gamma(1/4)}\,.
\eeq
 In order to account for expansion, we can indirectly exploit the conformal invariance of the model. Since $a \;\propto\; t^{1/2}$, conformal time is $\tau\equiv \int a^{-1}\,\diff t \propto a$, and $m_{\phi} t \;=\; m_{\rm end}(\tau-\tau_{\rm end})$. We can therefore characterize the stage of coherent oscillations of the inflaton by means of the approximate solution 
\beq\label{eq:phibackapp}
\phi(\tau) \;\simeq\; \phi_{\rm end}\left(\frac{a_{\rm end}}{a(\tau)} \right) {\rm sn}\,\left( \frac{m_{\rm end}}{\sqrt{6}}(\tau-\tau_{\rm end}),-1\right)\,.
\eeq

\subsection{Parametric resonance}
\label{sec:parametricresonance}

In the Introduction we have made the case for the study of the growth of the inflaton fluctuations during reheating, driven by the self-interaction. We now proceed to study the resonant growth of the inflaton momentum modes in the linear approximation. For these purposes, we denote the non-homogeneous inflaton perturbation as $\delta\phi(t,\bx)$. Variation of the action (\ref{eq:actionphi}), disregarding couplings of $\phi$ to other degrees of freedom, leads to the following dynamical equation,\footnote{In this work we ignore the fluctuations of the metric, which can potentially be important during the preheating stage~\cite{Nambu:1996gf,Bassett:1998wg,Bassett:1999mt,Jedamzik:2010dq,Huang:2011gf,Giblin:2019nuv}.}
\beq\label{eq:eomdeltaphi}
\ddot{\delta\phi} + 3H\dot{\delta\phi} - \frac{\nabla^2 \delta\phi}{a^2} + 12\lambda \phi(t)^2\,\delta\phi \;=\; 0\,.
\eeq
Introducing the canonically normalized fluctuation $X\equiv a\,\delta\phi$, and switching to conformal time, we can write the canonically quantized field as
\beq\label{eq:modedef}
X(\tau,\bx) = \int \frac{\diff  ^3\bk}{(2\pi)^{3/2}}\,e^{-i\bk\cdot\bx} \left[ X_k(\tau)\hat{a}_{\bk} + X_k^*(\tau)\hat{a}^{\dagger}_{-\bk} \right]\,, 
\eeq
where $\bk$ denotes the comoving momentum, and $\hat{a}_{\bk}$ and $\hat{a}^{\dagger}_{\bk}$ are the annihilation and creation operators, respectively, satisfying the canonical commutation relations $[\hat{a}_{\bk},\hat{a}^{\dagger}_{\bk'} ] = \delta(  \bk-\bk')$, $[\hat{a}_{\bk},\hat{a}_{\bk'} ] = [\hat{a}^{\dagger}_{\bk},\hat{a}^{\dagger}_{\bk'} ] = 0$. We ensure that the corresponding canonical commutation relations between the field, $X_k$, and its momentum conjugate, $X_k'$, are fulfilled by imposing the Wronskian constraint $X_kX^{*\prime}_k - X_k^*X_k' \;=\; i$. In this decomposition, the equation of motion (\ref{eq:eomdeltaphi}) reduces to
\beq\label{eq:eomX}
X_{k}'' + \left(k^2 - \frac{a''}{a} + 12\lambda \phi^2 a^2 \right) X_{k} \;=\; 0\,,
\eeq
where ${}^{\prime}$ denotes the derivative with respect to conformal time. Examination of the factor in parenthesis shows that the second term decreases over time; for it we have
\beq
\frac{a''}{a} \;=\; \frac{a^2}{6M_P^2} (4V-\dot{\phi}^2) \;\sim\; \frac{\lambda \phi^4 a^2}{M_P^2} \;\rightarrow\; 0\,.
\eeq
On the other hand, the third term inside the parenthesis of Eq.~(\ref{eq:eomX}) does not decrease with time (cf.~(\ref{eq:phiredshift})), and its time dependence is only modulated by $\mathcal{P}$. Introducing the dimensionless variable $z\equiv m_{\rm end}(\tau-\tau_{\rm end})$, we then find that soon after the onset of reheating, Eq.~(\ref{eq:eomX}) can be approximated as
\begin{figure}[!t]
\centering
    \includegraphics[width=0.82\textwidth]{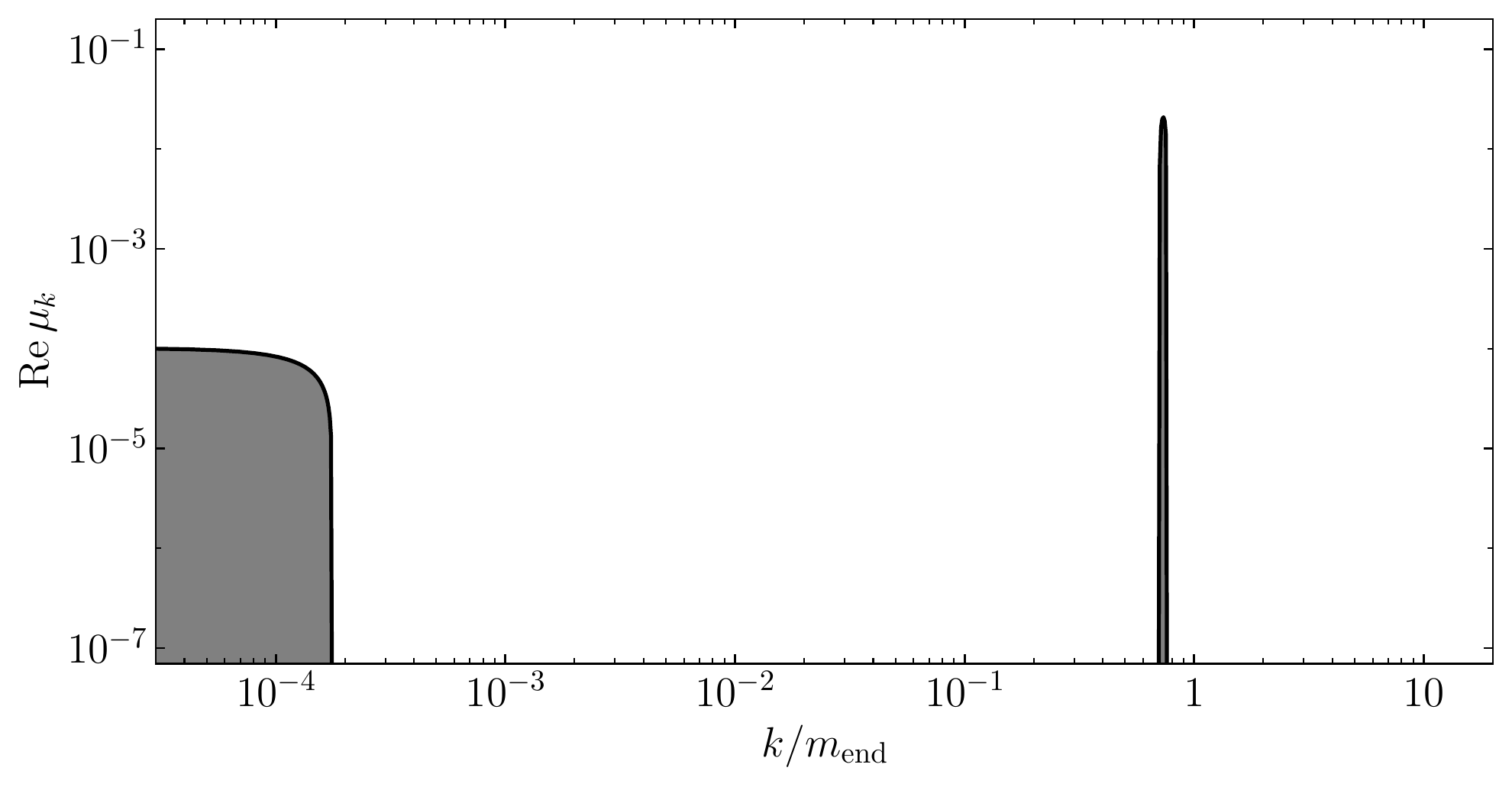}
    \caption{Floquet chart for Eq.~(\ref{eq:hill}). Due to the conformal nature of the model (\ref{eq:eomback}), a mode that begins inside a resonance band (shown in gray) stays indefinitely in the band, regardless of the expansion of the universe.}
    \label{fig:floquet}
\end{figure} 
\beq\label{eq:hill}
\dfrac{\diff^2 X_{k}}{\diff z^2} + \left[\left(\frac{k}{m_{\rm end}}\right)^2 + {\rm sn}^2\left( \frac{z}{\sqrt{6}},-1\right) \right]X_{k} \;=\;0\,.
\eeq
Eq.~(\ref{eq:hill}) has the form of Hill's equation (known as Mathieu's equation in the harmonic case), and presents parametric resonance. Floquet's theorem guarantees that the solutions to this equation have the form~\cite{Magnus2004-br}
\beq
X_{k}(z) \;=\; e^{\mu_k z}g_1(z) + e^{-\mu_k z}g_2(z)\,,
\eeq
where $g_1(z)$ and $g_2(z)$ are periodic functions, and $\mu_k$ is a complex number, called the Floquet exponent. Exponentially growing solutions are therefore found if ${\rm Re}\,\mu_k \;>\;0$. In order to determine the Floquet exponents, we follow the eigenvalue method described in detail in~\cite{Amin:2014eta}. The result of this numerical computation is shown in Fig.~\ref{fig:floquet}. The simplicity of the Floquet chart is noteworthy, as only two main resonance bands appear. One at small momentum values, $k/m_{\rm end}\lesssim 1.8\times 10^{-4}$, and a second one, with a narrow width and Floquet exponent more than two orders of magnitude larger, at $0.71 \lesssim k/m_{\rm end} \lesssim 0.76$. Due to the conformal nature of the quartic potential, the Floquet exponents are insensitive to the expansion of the universe, and a mode within a resonance band will stay in the band while the oscillation of the inflaton lasts. Therefore, modes with $k/m_{\rm end}\simeq 0.7$ will rapidly grow at the onset of reheating, and become non-linear after a few $e$-folds, even for a small coupling constant $\lambda$. 

\subsection{Backreaction and fragmentation}
\label{sec:backreaction}

\begin{figure}[!t]
\centering
    \includegraphics[width=\textwidth]{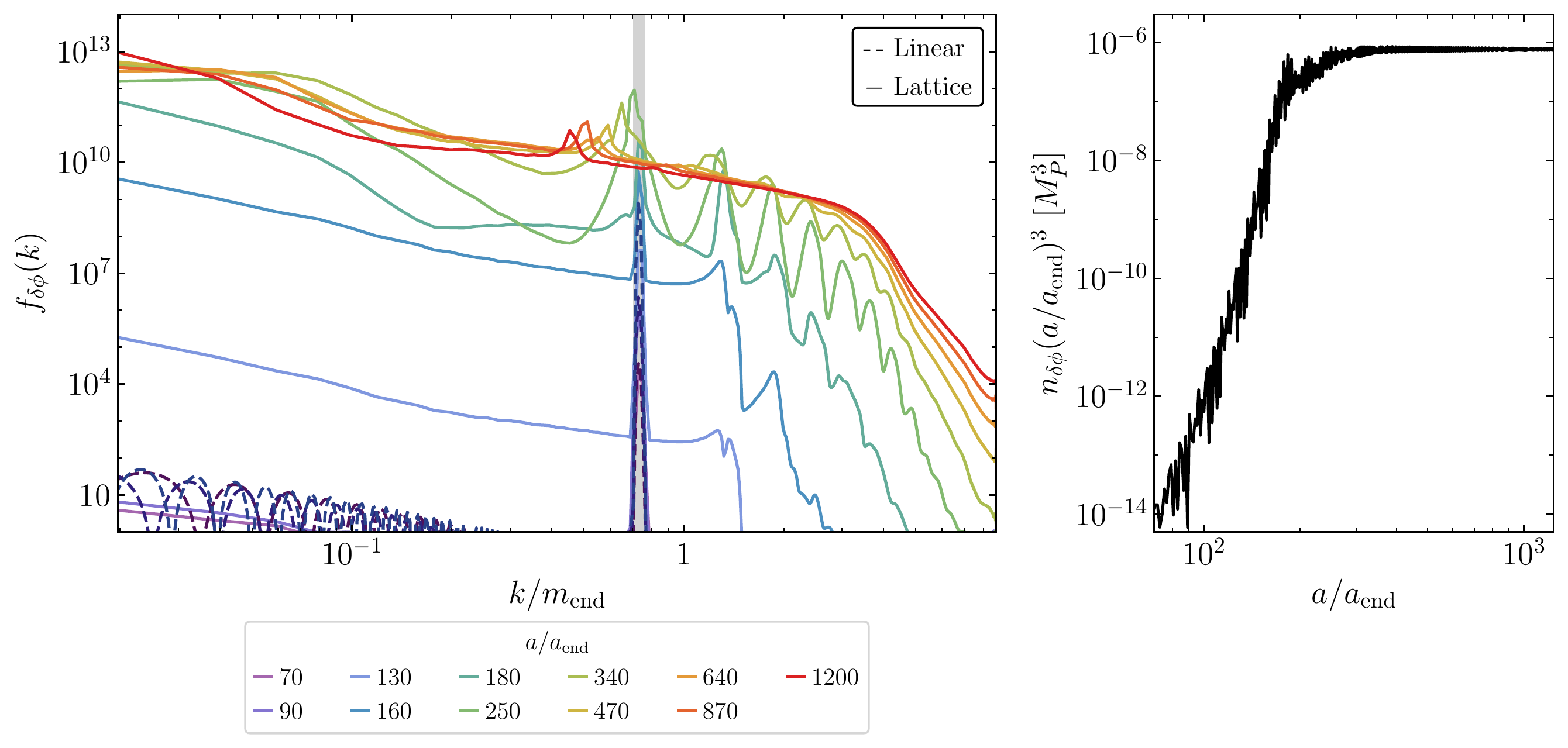}
    \caption{The particle phase space distribution (PSD) for the inflaton fluctuations for a selection of scale factors, coded by color (left), and the corresponding scale factor dependence of the comoving number density (right). The lattice results are shown in both panels as solid curves. In the left panel, only for the first three values of $a$ we also show the PSD as computed from the linear spectral analysis, as dashed lines. The main Floquet band is shown in light gray.}
    \label{fig:PSD}
\end{figure}

At linear order, the growth of the momentum modes of $X$ can be tracked by solving Eq.~(\ref{eq:eomX}) with the appropriate initial conditions, which we take as the positive-frequency Bunch-Davies vacuum
\beq
X_k(\tau_0) \;=\; \frac{1}{\sqrt{2\omega_k}} \,, \quad X_k'(\tau_0) \;=\; -\frac{i\omega_k}{\sqrt{2\omega_k}}\,,
\eeq
where 
\beq
\omega_k^2 \;=\; k^2 - \frac{a''}{a} + 12\lambda \phi^2 a^2\,.
\eeq 
Our main quantity of interest will be the phase space distribution (PSD) of the fluctuations, which coincides with the comoving occupation number of $\phi$. Its UV-finite form is given by~\cite{Kofman:1997yn}
\beq
f_{\delta\phi}(k, t)\;=\; n_{k} \;=\; \frac{1}{2\omega_k}\left| \omega_k X_k - i X'_k \right|^2\,.
\label{eq:PSDandoccupationnumber}
\eeq
Two other important quantities can be obtained from this PSD, the number density and the energy density of the fluctuations. Their UV-regular forms are computed as~\cite{Kofman:1997yn,Garcia:2021iag}
\begin{align}\label{eq:ndeltaphi}
n_{\delta\phi} \;&=\; \frac{1}{(2\pi)^3a^3}\int \diff^3\bk\, n_k\,,\\ \label{eq:rhodeltaphi}
\rho_{\delta\phi} \;&=\; \frac{1}{(2\pi)^3a^4}\int \diff^3\bk\, \omega_k n_k\,.
\end{align}
The resulting form of the PSD in the linear approximation is shown as the dashed lines in the left panel of Fig.~\ref{fig:PSD}. As expected, the PSD is peaked around $k/m_{\rm end}=0.73$, in full agreement with the results of the Floquet analysis of the previous section. The height of this peak increases with time, as the resonant growth accumulates during the oscillation of $\phi$. We only show three snapshots of the PSD for $a\leq 130$, since the system rapidly evolves toward the non-linear regime, making the analysis based on the solution of (\ref{eq:eomX}) unsuitable. 

\begin{figure}[!t]
\centering
    \includegraphics[width=0.9\textwidth]{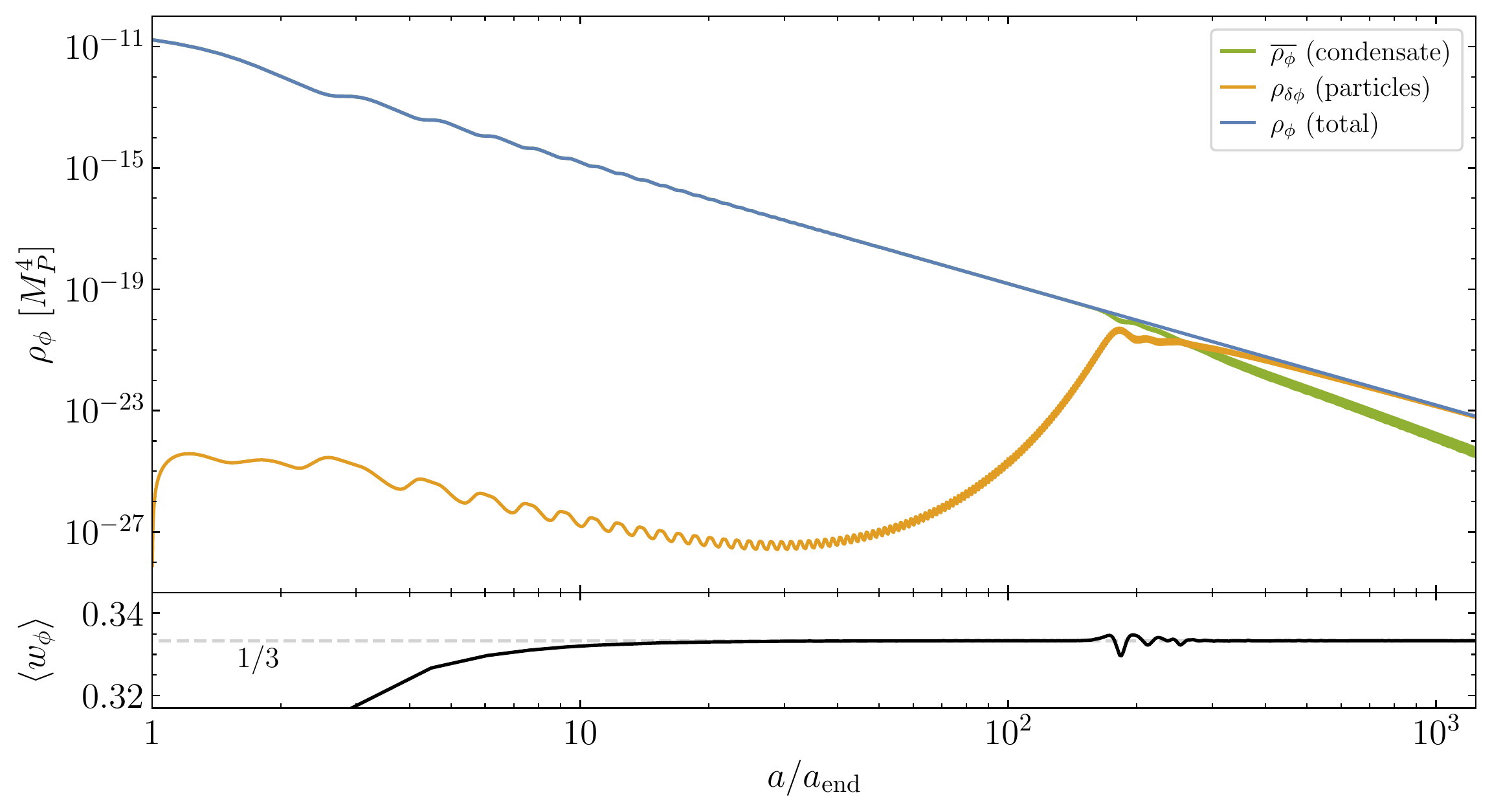}
    \caption{Top: Pre-~and post-fragmentation evolution of the total inflaton energy density (blue), the energy density of the inhomogeneous component of $\phi$ (orange) and the energy density of the homogeneous component (green). For $\rho_{\delta\phi}$, the linear approximation (\ref{eq:rhodeltaphi}) is depicted up to $a/a_{\rm end}\leq 10^2$. At later times the lattice result $\rho_{\delta\phi}=\rho_{\phi}-\overbar{\rho_{\phi}}$ is shown. Bottom: Oscillation-averaged inflaton equation of state parameter.}
    \label{fig:rhos}
\end{figure}

The impact of the rapid growth of the resonant modes on the energy density of the inflaton fluctuations can also be appreciated in Fig.~\ref{fig:rhos}. We follow the scale factor dependence of $\rho_{\delta\phi}$ (the orange curve) by means of (\ref{eq:rhodeltaphi}) only up to $a/a_{\rm end}=10^2$. Within this range, the growth of the energy density in fluctuations is clear, ramping up at $a/a_{\rm end}\sim 30$, and growing exponentially fast afterwards. The transition from the linear to the non-linear regime is clear.\par\medskip

From this discussion it follows that, despite the smallness of $\lambda$, the growth of the momentum modes in the main resonance band is strong enough to eventually be able to draw an $\mathcal{O}(1)$ fraction of the energy density of the zero-mode of $\phi$. The inflaton ``fluctuation'' is no longer such, and the system enters the non-linear regime. Mode-mode couplings become important, redistributing the energy of the resonant mode into other modes ({\em rescattering}), leading to large configuration space gradients,  and effectively fragmenting the once homogeneous condensate. This phenomenon is known as {\em backreaction}, and it does not allow for a straightforward spectral analysis. Instead, the full non-linear configuration-space equation of motion for the full operator $\phi(t,\bx)$ must be solved, which is also a non-trivial task. To get around some of the complications, it is argued that, since in the non-linear regime occupation numbers are large, $n_k\gg 1$, the dynamics of the system can be adequately tracked by approximating $\phi$ as a classical field. The full non-linear PDE
\beq
\ddot{\phi} + 3H\dot{\phi} - \frac{\nabla^2\phi}{a^2} + V_\phi \;=\; 0\,,
\eeq
is then solved over a configuration-space lattice. The energy density of the inflaton is computed from the spatial average of the energy-momentum tensor of $\phi$, which we denote with an over-bar,
\beq
\rho_{\phi} \;=\; \overbar{\frac{1}{2}\dot{\phi}^2 + \frac{1}{2a^2}(\nabla\phi)^2+V(\phi)}\,.
\eeq
Spectral data in turn is obtained upon Fourier transformation of configuration-space quantities. The quartic inflaton system has been extensively studied in the past by means of lattice methods, see e.g.~\cite{Frolov:2010sz,Figueroa:2016wxr,Lozanov:2016hid,Lozanov:2017hjm,Fu:2017ero,Antusch:2021aiw,Lebedev:2022vwf,Alam:2023kia}. For our analysis we use the publicly available code \CL~\cite{Figueroa:2020rrl,Figueroa:2021yhd}.

Our lattice results are summarized in Figs.~\ref{fig:PSD} and \ref{fig:rhos}. The left panel of Fig.~\ref{fig:PSD} shows, as continuous curves, the form of the PSD of the inflaton as computed by the lattice code. At early times it shares the features of the linear approximation, showing a main growing peak, located at the momentum band predicted by the Floquet analysis. For the earliest times, the lattice PSD coincides with the linear prediction. Nevertheless, for $a/a_{\rm end}\gtrsim 10^2$, the effect of rescattering becomes evident. The growth of the resonant mode is traded for the population of the PSD for lower momentum modes, and larger momentum adjacent modes. As time passes, the redistribution of energy becomes more efficient, and a UV tail in the distribution appears. At first this tail is populated more efficiently at some discrete values of $k$. Notably, these wavenumbers can be estimated by means of the Boltzmann approximation, as we show in Section~\ref{sec:boltzmann}. In any case, these features are quickly erased, and a smooth limiting distribution emerges, shown in red. The right panel of Fig.~\ref{fig:PSD} conveys this asymptotic behavior of the PSD, in terms of the comoving number density of $\phi$. At first $n_{\delta\phi}$ grows rapidly, driven by the parametric resonance. Notice that for $a/a_\text{end}>180$, the resonance peak in the occupation number progressively shifts towards lower scales. The subsistence of the inflaton condensate after fragmentation implies the subsistence of a resonance. However, as the inflaton condensate is more strongly affected by redshift after fragmentation, scales sensitive to parametric resonance also experience a redshift in a similar manner, explaining the progressive shift of the peak from $k/m_\text{end}\simeq 0.7$ (blue curve) to $k/m_\text{end}\simeq 0.4$ (red curve) observed in Fig.~\ref{fig:PSD}. For $a/a_{\rm end}\gtrsim 300$, the comoving number density freezes, and the effect of rescatterings is merely the kinetic redistribution of energy among modes.

Fig.~\ref{fig:rhos} contains the lattice results for the total energy density of $\phi$, the density of the condensate, and that of free particles, for $a/a_{\rm end}>10^2$. For definiteness, and in order to connect with the results of the spectral analysis, we define the condensate component of the energy density as follows~\cite{Garcia:2021iag},
\beq
\overline{\rho_{\phi}} \;=\; \frac{1}{2}\overbar{\dot{\phi}}^2 + V(\overbar{\phi})\,,
\eeq
that is, the energy density of the spatially averaged inflaton field. In turn, we take $\rho_{\delta\phi}=\rho_{\phi}-\overbar{\rho_{\phi}}$. This definition of the fluctuation energy density matches excellently with the spectral result, and is shown as the orange curve in Fig.~\ref{fig:rhos}. Its value rapidly grows until $a/a_{\rm end}\simeq 180$, point at which the backreaction in the oscillating condensate is noticeable. The total energy density of $\phi$ redshifts as radiation even after rescattering becomes important, but the energy density of the zero mode is now efficiently transferred to fluctuations. Interestingly, despite the initial rapid growth in $\rho_{\delta\phi}$, the dissipation of $\overbar{\rho_{\phi}}$ after entering backreaction is gradual, $\overbar{\rho_{\phi}}\propto a^{-5.3}$, unlike the exponential decrease of preheating scenarios in quadratic potentials~\cite{Garcia:2021iag}. The survival of this coherent component will be an important ingredient in our exploration of the decay of $\phi$ of Section~\ref{sec:reheating}.

\subsection{The Boltzmann approximation}\label{sec:boltzmann}

In the previous sections we have shown that a combination of spectral and lattice methods is necessary to correctly determine the distribution of the inflaton fluctuations. We now take the opportunity to show that, in the linear regime, it is possible to extract non-trivial information for the PSD by means of the integration of the Boltzmann equation for the inflaton fluctuations. Inflaton quanta $\delta \phi$ sourced from the inflaton background $\phi$ are induced by the interaction Lagrangian\footnote{Additional terms such as $\phi \delta \phi^3$ are also present but the corresponding production rate would be suppressed by phase-space volume therefore disregarded here.}
\begin{equation}
\mathcal{L}_{I}=6 \lambda \phi^{2} \delta \phi^{2}\,.
\label{eq:LI}
\end{equation}
Following~\cite{Nurmi:2015ema,Garcia:2020wiy,Garcia:2022vwm}, the Boltzmann equation in the presence of anharmonic oscillations of the inflaton background takes the form 

\begin{equation}
\begin{split}
   \frac{\partial f_{\delta \phi}}{\partial t}-H|\boldsymbol{P}| \frac{\partial f_{\delta \phi}}{\partial|\boldsymbol{P}|}\,=\,\frac{1}{P^{0}} &\sum_{n=1}^{\infty} \int  \frac{\diff^{3} \boldsymbol{K}_{n}}{(2 \pi)^{3} n_{\phi}} \frac{\diff^{3} \boldsymbol{P}^{\prime}}{(2 \pi)^{3} 2 P^{\prime 0}}(2 \pi)^{4} \delta^{(4)}\left(K_{n}-P-P^{\prime}\right)\left|\overbar{\mathcal{M}_{n}}\right|^{2} \\ & \times\left[f_{\phi}\left(K_{n}\right)\left(1+f_{\delta \phi}(P)\right)\left(1+f_{\delta \phi}\left(P^{\prime}\right)\right)-f_{\delta \phi}(P) f_{\delta \phi}\left(P^{\prime}\right)\left(1+f_{\phi}\left(K_{n}\right)\right)\right] . 
\label{eq:Boltzmannfordeltaphi}
\end{split}
\end{equation}
Here we denote the physical four-momenta of the fluctuations by $P$ and $P'$. $K_n = \left(E_n, \bf{0} \right)$ is the physical four-momentum of the inflaton condensate, where $E_n = n \, \omega_{\phi}$ denotes the energy of the $n^\text{th}$ oscillating mode.  $\mathcal{M}_{n}$ represents the transition amplitude corresponding to  the production of a pair of inflaton quanta $| f\rangle=| \delta \phi \delta \phi\rangle$ from the $n^\text{th}$ Fourier mode of the coherently oscillating inflaton background field.
\begin{equation}
\big|\big\langle f \big| i \int \diff^{4} x \, \mathcal{L}_{I} \big| 0 \big\rangle \big|^{2}=\operatorname{Vol}_{4} \sum_{n=-\infty}^{\infty}\left|\overline{\mathcal{M}_{n}}\right|^{2}(2 \pi)^{4} \delta^{(4)}\left(K_{n}-P-P^{\prime}\right) .
\label{eq:matrixelement}
\end{equation}
where Vol$_4$ denote the space-time volume. From Eq.~(\ref{eq:LI}), the matrix elements $\mathcal{M}_{n}$ are found to be
\begin{equation}
\mathcal{M}_{n}\;=\;12 \lambda \phi_{0}^{2}\,\hat{\mathcal{P}}_{n} \qquad \Rightarrow \qquad\left|\overline{\mathcal{M}_{n}}\right|^{2}=72 \lambda^{2} \phi_{0}^{4}\,|\hat{\mathcal{P}}_{n}|^{2}\,.
\end{equation}
The mean squared amplitude $|\overline{\mathcal{M}_{n}}|^2$ accounts for a factor of 2 for identical particles in the final state. The $\hat{\mathcal{P}}_n$ are the Fourier coefficients of the harmonic decomposition over one oscillation of the square of the quasi-periodic function $\mathcal{P}(t)$ defined in (\ref{eq:pdef}),
\begin{equation}
\mathcal{P}^2(t) \;=\; \sum_{n=-\infty}^{\infty} \hat{\mathcal{P}}_n e^{-in\omega_{\phi} t} \, .
\end{equation}
The spatially homogeneous background inflaton condensate distribution can be expressed as the zero-mode
\begin{equation}
f_{\phi}(\boldsymbol{K}, t)=(2 \pi)^{3} n_{\phi}(t) \delta^{(3)}(\boldsymbol{K})\,.
\end{equation}
Neglecting the backreaction of the inflaton quanta onto the condensate, the Boltzmann equation takes the following form
\begin{align} \notag
\frac{\partial f_{\delta \phi}}{\partial t}&\;-\;H|\boldsymbol{P}| \frac{\partial f_{\delta \phi}}{\partial|\boldsymbol{P}|}  \\ \displaybreak[0]
&\simeq \; \sum_{n=1}^{\infty} \frac{1}{P^{0}} \int \frac{\diff^{3} \boldsymbol{K}_{n}}{(2 \pi)^{3} n_{\phi}} \frac{\diff^{3} \boldsymbol{P}^{\prime}}{(2 \pi)^{3} 2 P^{\prime 0}}(2 \pi)^{4} \delta^{(4)}\left(K_{n}-P-P^{\prime}\right)\left|\overline{\mathcal{M}_{n}}\right|^{2} f_{\phi}\left(K_{n}\right)\left(1+f_{\delta \phi}(P)+f_{\delta \phi}\left(P^{\prime}\right)\right) \nonumber\\ \notag
& \simeq\; 72 \pi \lambda^{2} \phi_{0}^{4} \sum_{n=1}^{\infty} \int \frac{\diff^{3} \boldsymbol{P}^{\prime}}{P^{0} P^{\prime 0}} \delta\left(n \omega_\phi-P^{0}-P^{\prime 0}\right) \delta^{(3)}\left(\boldsymbol{P}+\boldsymbol{P}^{\prime}\right) \, |\hat{\mathcal{P}}_n|^{2} \left(1+f_{\delta \phi}(P)+f_{\delta \phi}\left(P^{\prime}\right)\right) \\
& = \; 144 \pi \lambda^{2} \phi_{0}^{4} \sum_{n=1}^{\infty} \frac{|\hat{\mathcal{P}}_n|^{2}}{n^{2} \omega_\phi^{2} \beta_{n}} \delta\left(|\boldsymbol{P}|-\frac{1}{2} n \omega_\phi \beta_{n}\right)\Big(1+2 f_{\delta \phi}(|\boldsymbol{P}|)\Big)\,,
\label{eq:Boltzmannsimp}
\end{align}
where
\begin{equation}
\beta_{n}\;=\;\sqrt{1-\frac{4 m_{\phi}^{2}}{n^{2} \omega_\phi^{2}}} \;=\; \sqrt{1-\left(\frac{2}{n c}\right)^{2}}, \quad  \quad c \;\equiv\; \sqrt{\frac{2 \pi}{3}} \frac{\Gamma(3 / 4)}{\Gamma(1 / 4)}\,,
\end{equation}
denotes the (time-independent) kinematic suppression factor. The quantum mechanical Bose enhancement factor, which appears in the right side of Eq.~(\ref{eq:Boltzmannsimp}), can be scaled out by defining the ``classical" distribution $f_{\chi}^c$~\cite{Moroi:2020has, Moroi:2020bkq, Ghosh:2022hen}:
\beq\label{eq:boltzfexp}
f_{\delta \phi}(|\boldsymbol{P}|, t) \;\equiv\; \frac{1}{2}\Big[ \exp\big(2f_{\delta \phi}^c(|\boldsymbol{P}|,t)\big) - 1\Big]\,.
\eeq
The Boltzmann equation reduces then to
\begin{equation}
\begin{aligned}
\frac{\partial f^c_{\delta \phi}}{\partial t}-H|\boldsymbol{P}| \frac{\partial f^c_{\delta \phi}}{\partial|\boldsymbol{P}|}  \;=\; 144 \pi \lambda^{2} \phi_{0}^{4} \sum_{n=1}^{\infty} \frac{|\hat{\mathcal{P}}_n|^{2}}{n^{2} \omega_\phi^{2} \beta_{n}} \delta\left(|\boldsymbol{P}|-\frac{1}{2} n \omega_\phi \beta_{n}\right)\,.
\end{aligned}
\label{eq:Boltzmannsimp2}
\end{equation}
As the right-hand side of Eq.~(\ref{eq:Boltzmannsimp2}) is $f^c_{\delta \phi}$-independent, we can immediately integrate this equation to get~\cite{Garcia:2018wtq,Ballesteros:2020adh}
\begin{equation}
\begin{aligned}
f^c_{\delta \phi}(|\boldsymbol{P}|, t) \;& \simeq \;144 \pi \lambda^{2} \sum_{n=1}^{\infty} \frac{|\hat{\mathcal{P}}_n|^{2}}{n^{2} c^{2} \beta_{n}} \int_{t_{\mathrm{end}}}^{t} \diff t^{\prime} \frac{\phi_{0}^{4}\left(t^{\prime}\right)}{m_{\phi}^{2}\left(t^{\prime}\right)} \delta\left(\frac{a(t)}{a\left(t^{\prime}\right)}|\boldsymbol{P}|-\frac{1}{2} n c \beta_{n} m_{\phi}(t^{\prime})\right) \\
& \simeq\; \frac{\pi}{c^{2}}\left(\frac{m_{\mathrm{end}}}{H_{\mathrm{end}}}\right)\left(\frac{a(t)}{a_{\mathrm{end}}}-1\right) \sum_{n=1}^{\infty} \frac{|\hat{\mathcal{P}}_n|^{2}}{n^{2} \beta_{n}} \delta\left(q-\frac{1}{2} n c \beta_{n}\right)\,, 
\end{aligned}
\end{equation}
with $q\equiv a(t) |\boldsymbol{P}|/(a_\text{end} m_\text{end})$ the re-scaled comoving momentum referred to the end of inflation, denoted as $k/m_{\rm end}$ in Fig.~\ref{fig:PSD}. \par \medskip

\noindent
\textbf{Peak structure.} 
The distribution function corresponds to a collection of Dirac delta functions, located at different momenta $(1/2)n c \beta_{n}$. The $\beta_n$ coefficient is a decreasing function of $n$ whose first (real) non-vanishing value is achieved for $n=5$. By denoting the location of the $i^\text{th}$ peak as $\hat{q}_i$, the distribution can be equivalently expressed as as sum of contributing modes via
\begin{equation}
f^c_{\delta \phi}(q, t) \, = \, \sum_{i=1}^{\infty} \mathcal{F}^c_i (t) \, \delta\left(q-\hat{q}_i\right)\,,
\end{equation}
with time-dependent coefficients $\mathcal{F}^c_i (t)$ and peak locations determined by
\begin{equation}
    \hat{q}_i \, = \, \sqrt{\frac{2 \pi }{3}} \, \left(\dfrac{i+4}{2} \right)  \dfrac{\Gamma(3/4)}{\Gamma(1/4)} \sqrt{1-\left(\frac{6 }{\pi (i+4)^2 } \right) \left(\dfrac{\Gamma(1/4)}{\Gamma(3/4)} \right)^2}\,,
    \label{eq:Boltzmannpeak}
\end{equation}
where $i=n-4$. This ratio is remarkably independent of any model parameter. It is straightforward to check that the first peak $\hat{q}_1$ is predicted to be at $\hat{q}_1\simeq 0.7$, in excellent agreement with the Floquet analysis and the linear and lattice results. The relative location of peaks at larger momenta can be straightforwardly inferred from Eq.~(\ref{eq:Boltzmannpeak}) and corresponds to the following ratio
\begin{equation}
\dfrac{\hat q_i}{\hat q_j} \, = \, \dfrac{(i+4) \beta_{i+4}}{(j+4) \beta_{j+4}}\,.
\label{eq:Boltzmannpeakprediction}
\end{equation}
\par \medskip

\noindent
\textbf{Limitation of the Boltzmann approach.} The previous computation resulted in a distribution peaked at discrete values of momenta, with an infinitesimally narrow width and a large amplitude peak characteristic of a Dirac delta function. In reality however, as we have seen in the previous sections, the distribution is spread around the peaks. Immediately after the end of inflation, from our linear approach we found a distribution function at the first peak $f_{\delta \phi}(\hat q_1,t_\text{end}) \sim 0.2$. Moreover, the mode amplitudes are expected to grow continuously $\mathcal{F}^c_i (t) \propto a(t)/a_\text{end}$ during reheating, becoming rapidly even larger. For such large amplitudes the Boltzmann approach breaks down almost immediately~\cite{Moroi:2020bkq,Garcia:2022vwm}.  \par \medskip

The amplitude of the peaks, therefore, cannot be accurately estimated from the Boltzmann approach starting from the first instants after the end of inflation. However, we find that this approach gives a qualitatively good description of the location of the PSD peaks appearing at the onset of backreaction, and notably, for the peaks in the induced gravitational wave spectrum. We postpone the detailed comparison to Section~\ref{sec:gws}. \par \medskip

\section{Reheating in a quartic potential}\label{sec:reheating}

We now turn to the study of the decay of the inflaton into light degrees of freedom, necessary to complete reheating and populate the universe with a relativistic plasma in thermal equilibrium. As mentioned in the Introduction, our analysis will be based on the assumption that fields coupled to the inflaton are not strongly sourced via parametric resonance. To satisfy this assumption, we will implicitly assume that the main decay products of $\phi$ are spin 1/2 fermions, denoted as $\psi$. Moreover, we will assume for simplicity that the decay of the inflaton occurs at tree-level, and produces two particles in the final state. Thus, generically, we consider
\beq\label{eq:Lintferm}
\mathcal{L}_{\rm int} \;=\;  g\phi\bar{\psi}\psi + i g'\phi\bar{\psi}\gamma_5\psi\,.
\eeq
It is worth noting that the Yukawa term in the previous expression can lead to fermion preheating~\cite{Greene:1998nh,Giudice:1999fb,Greene:2000ew,Peloso:2000hy,Nilles:2001fg}, for which the production of particles is resonantly {\em suppressed}, a consequence of Fermi-Dirac statistics. Perturbatively, the oscillating $\phi$ induces an effective mass for $\psi$ than can kinematically block the decay into $\psi$ very efficiently~\cite{Garcia:2020wiy}. The exploration of these interesting effects is not the main purpose of the present work, and will be postponed for a future study. We will therefore disregard these mechanisms, a possibility that arises for $g'\gg g$.

\subsection{Production rates}
\label{sec:rates}

Schematically, the Boltzmann equation that must be integrated to track the phase space distribution of the decay products can be written as
\beq\label{eq:genfR}
\frac{\partial f_{\psi}}{\partial t} - H|\boldsymbol{P}|\frac{\partial f_{\psi}}{\partial |\boldsymbol{P}|} \;=\; \mathcal{C}[f_{\psi}]\,,
\eeq
with $\mathcal{C}$ denoting the collision term. For generality (and future applications) let us for now assume a non-vanishing, and potentially time-dependent mass $m_{\psi}(t)$ for the decay products. Applying the operator $\int \diff^3 \boldsymbol{P} \, P^0/(2\pi)^3 \,$ on both sides of (\ref{eq:genfR}), we obtain the following general form for the continuity equation for the energy density $\rho_{\psi}$,
\beq
\dot{\rho}_{\psi} + 4H\rho_{\psi} - \int \frac{\diff^3 \boldsymbol{P} }{(2\pi)^3P^0}\, f_{\psi}(P)\,m_{\psi}\left(\dot{m}_{\psi} + H m_{\psi}\right) \;=\; \int \frac{\diff^3 \boldsymbol{P} }{(2\pi)^3}\,P^0 \mathcal{C}[f_{\psi}] \;\equiv\; R(t)\,.
\label{eq:continuityradiation}
\eeq
Here we introduce $R(t)$, the radiation-energy production rate per unit of volume. 

In the following we assume that the states $\psi$ eventually thermalize among themselves and the rest of the Standard Model states. We therefore identify $\rho_{\psi}=\rho_R$, where the later denotes the energy density of the relativistic radiation plasma. The evaluation of the production rate in Eq.~(\ref{eq:continuityradiation}) will be split in two. We estimate the contributions to the production rate from the oscillating inflaton condensate $\phi$ and from the fragmentated inflaton quanta $\delta \phi$
\begin{equation}
    R(t) \, \equiv \, R_\phi(t)+R_{\delta \phi}(t) \,.
    \label{eq:totalrate}
\end{equation}

\subsubsection{Decay of the coherent oscillations}\label{sec:coherentdecay}

Let us first briefly discuss the decay of the inflaton at early times, when it is composed mainly of the coherent, oscillating zero-mode. Under the assumption that quantum statistics do not play a role in the dissipation process, the Boltzmann equation which describes the decay of the inflaton can be written as~\cite{Kainulainen:2016vzv,Ichikawa:2008ne,Garcia:2020wiy}
\beq\label{eq:Boltzcoherent}
\dot{\rho}_{\phi} + 3H(1+w_{\phi})\rho_{\phi} \;=\;  -R_\phi(t)\,.
\eeq 
The rate in this equation is evaluated as
\begin{equation}
R_\phi(t) \, = \,  (1+w_{\phi})\Gamma_{\phi} \rho_{\phi}\,,
\end{equation}
where the decay rate $\Gamma_\phi$ for a generic process $\phi\rightarrow A+B$ is
\begin{equation}
\Gamma_{\phi} \;=\; \frac{1}{8\pi(1+w_{\phi})\rho_{\phi}}\sum_{n=1}^{\infty}|\mathcal{M}_n|^2E_n\sqrt{\left(1-\frac{(m_A+m_B)^2}{E_n^2}\right)\left(1-\frac{(m_A-m_B)^2}{E_n^2}\right)}\,.
\end{equation}
Here the sum is taken over the harmonic modes of the function $\mathcal{P}(t)$ over one oscillation, with associated energy $E_n=n\omega_{\phi}$ (see Eq.~(\ref{eq:expansionP})). $\mathcal{M}_n$ denotes the transition amplitude in one oscillation for each mode from the coherent state to the two-particle state that can be defined analogously to Eq.~(\ref{eq:matrixelement}), with $|f\rangle=|A,B\rangle$ and $\mathcal{L}_I=\mathcal{L}_\text{int}$, cf.~(\ref{eq:actionphi}). In the case of an inflaton oscillating about a quartic potential, assuming a fermionic decay for which the masses can be disregarded, the decay rate $\Gamma_{\phi} $ can be evaluated explicitly, taking the form~\cite{Garcia:2020wiy}
\beq\label{eq:Gammacond}
\Gamma_{\phi\rightarrow\bar{\psi}\psi}(t)\;=\; \alpha^2 \, \frac{y^2}{8\pi}m_{\phi}(t)\,,
\eeq
where $y=\{g,g'\}$, $m_{\phi}$ is the envelope inflaton mass (\ref{eq:mphi}), and $\alpha\simeq 0.71$ is an efficiency factor that codifies the anharmonicity of the $\phi$ oscillation~\cite{Garcia:2020wiy}. The continuity-Friedmann system that completes Eq.~(\ref{eq:Boltzcoherent}) corresponds to
\begin{align}\label{eq:BoltzRcond}
\dot{\rho}_R + 4H\rho_R \;&=\; (1+w_{\phi})\,\Gamma_{\phi}\,\rho_{\phi}\,,\\
\rho_{\phi} + \rho_{R} \;&=\; 3M_P^2H^2\,,
\end{align}
with $w_{\phi}=1/3$. 

\subsubsection{Decay of the fragmented inflaton}\label{sec:fragdecay}

In Section~\ref{sec:backreaction} we discuss how the inflaton is fragmented after it enters the backreaction regime. When this occurs, the production of particles can no longer be characterized by (\ref{eq:BoltzRcond}) with decay rate (\ref{eq:Gammacond}). Instead, one must now determine the form of the continuity equation for $\rho_R$ from the microscopic Boltzmann equation, assuming a population of inflaton particles with PSD $f_{\delta\phi}(K)$. Disregarding inverse decays and Pauli blocking / Bose enhancement, the collision term in the Boltzmann equation (\ref{eq:genfR}) has the form
\begin{align} \label{eq:Boltzmicro}
\mathcal{C}_{\delta\phi}[f_{\psi}] \;=\; \frac{1}{P^0} \int &\frac{\diff^3 \boldsymbol{K} }{(2\pi)^3 2K^0} \frac{\diff^3 \boldsymbol{P}'}{(2\pi)^3 2P^{0\prime}}(2\pi)^4 \delta^{(4)}(K-P-P') |\mathcal{M}_{\delta \phi \rightarrow \bar \psi \psi}|^2 f_{\delta \phi}(K)\,.
\end{align}
The particle production rate can then be evaluated in a straightforward way, and yields
\beq
R_{\delta \phi}(t) \;=\; \int \frac{\diff^3 \boldsymbol{P} }{(2\pi)^3}\,P^0\, \mathcal{C}_{\delta\phi}[f_{\psi}] \;=\; \Gamma_{\delta \phi} m_{\phi}n_{\delta\phi}\,,
\eeq
where
\beq
\Gamma_{\delta \phi} \;=\; \frac{|\mathcal{M}_{\delta \phi \rightarrow \bar \psi \psi}|^2}{16\pi m_{\phi}}\sqrt{1-\frac{4m_{\psi}^2}{m_{\phi}^2}}\,,
\eeq
denotes the decay rate of free inflaton {\em particles} $\delta \phi$. In the case corresponding to (\ref{eq:Lintferm}), $\Gamma_{\delta \phi\rightarrow\bar{\psi}\psi}=y^2m_{\phi}/8\pi$ for negligible masses for the outgoing states. 

\subsubsection{Total rate}

Summarizing our previous findings, the total rate (\ref{eq:totalrate}) accounting for production from the inflaton condensate and particle rates reads
\begin{align}
R(t) \;&=\; \frac{4}{3}\Gamma_{\phi} \overline{\rho_{\phi}} + \Gamma_{\delta \phi} m_{\phi}n_{\delta\phi}\\
&\;=\; \frac{y^2}{8\pi}m_{\phi}\left(\frac{4}{3}\alpha^2\overline{\rho_{\phi}} + m_{\phi}n_{\delta\phi}\right)\,.
\end{align}
The second line corresponds to the two-body fermionic decay channel, in the limit where $m_{\psi}\ll m_{\phi}$.

\begin{figure}[!t]
\centering
    \includegraphics[width=\textwidth]{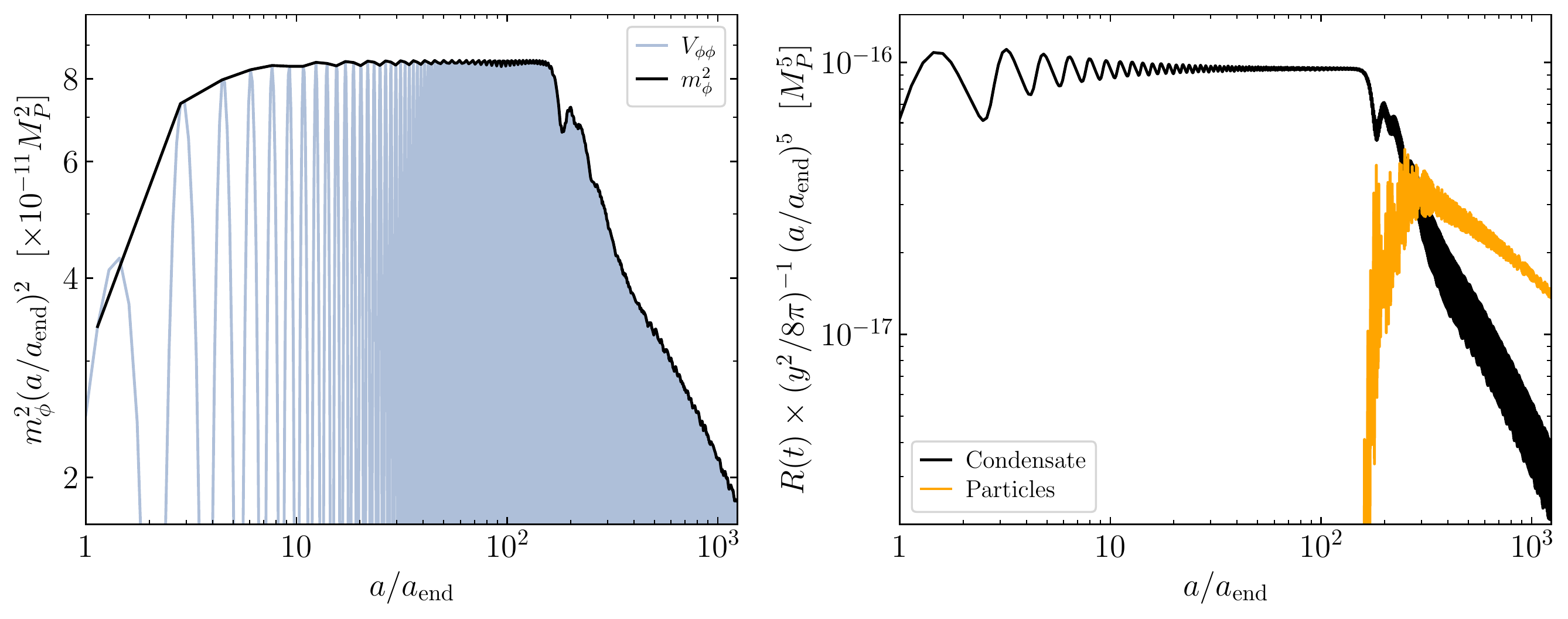}
    \caption{Scale factor dependence of the effective inflaton mass (left), and of the effective Boltzmann rate for the relativistic decay products $R(t)$ (right). The total rate is the sum of the condensate contribution (black) and the particle contribution (orange).}
    \label{fig:rates}
\end{figure} 

Fig.~\ref{fig:rates} shows the scale factor dependence of the effective mass of the inflaton, inherited from its coherent oscillation (left), and that of the rate $R(t)$ decomposed into the condensate ($R_\phi(t)$) and particle  ($R_{\delta \phi}(t)$) contributions (right). In both panels we have multiplied the corresponding variable by its redshift factor prior to $\phi$ fragmentation. For the effective mass, we have $m_{\phi}(a/a_{\rm end})\simeq {\rm const.}$ quickly after the beginning of reheating, until $a/a_{\rm end} \simeq 160$, at the onset of strong rescattering. Due to the efficient conversion of the inflaton zero-mode into finite momentum particles, $m_{\phi}$ is further reduced, first in a non-monotonic manner while strong backreaction takes place, and monotonically for $a\gtrsim 300 \,a_{\rm end}$, with $m_{\phi}\propto (a/a_{\rm end})^{1.32}$, as determined from the lattice data.

In the right panel of Fig.~\ref{fig:rates} we show the condensate (black) and particle (orange) contributions to the total rate $R(t)$, rescaled by its scale factor dependence and by the amplitude parameter $y^2/8\pi$. Similarly to the effective mass, for $a/a_{\rm end}\lesssim 160$, the rate is solely determined by the anharmonic oscillation of the inflaton, $R(t)\propto m_{\phi}\rho_{\phi}\propto (a/a_{\rm end})^{-5}$. However, at later times, the parametric growth of non-vanishing $k$ modes seeps energy from the classical $\phi$, reducing accordingly the corresponding rate. For the condensate component, at late times, we find $R_\phi \propto m_{\phi}\overbar{\rho_{\phi}}\propto (a/a_\text{end})^{-6.6}$. On the other hand, following the orange curve, we see that the contribution to $R$ from the fragmented inflaton becomes quickly important, and in fact dominates the dissipation process for $a/a_{\rm end}\gtrsim 200$. Since at this stage the comoving number density of $\delta\phi$ is conserved, we have $R_{\delta \phi}\propto m_{\phi}^2n_{\delta\phi}\propto a^{-5.65}$. 

\subsection{Reheating temperatures}\label{eq:rehtemp}

With the Boltzmann rate $R(t)$ at hand, we can now estimate the effect of fragmentation on the instantaneous temperature of the primordial plasma during reheating, and its value at the beginning of the domination by the thermal bath, which we denote by $T_{\rm reh}$. For simplicity we assume the instantaneous thermalization of the inflaton decay products, and for definiteness we assume an effective number of degrees of freedom $g_{\rm reh}=427/4$ above the electroweak scale, which is the Standard Model value.

\subsubsection{Condensate decay}

If we neglect the effect of backreaction, or assume a large coupling constant so that the decay occurs prior to fragmentation, we can use the condensate form for $R(t)$. Fixing for simplicity $a_{\rm end}=1$ for now, Eq.~(\ref{eq:continuityradiation}) can be written as
\beq\label{eq:boltzina}
\frac{\diff}{\diff a}\left(\rho_R a^4\right) \;=\; \frac{R(a)\,a^3}{H}\,,
\eeq
in general, and as
\beq
\frac{\diff}{\diff a}\left(\rho_R a^4\right) \;=\; \frac{\alpha^2y^2\lambda^{1/4}(\rho_{\phi}a^4)^{5/4}M_P}{\pi\sqrt{(\rho_{\phi}+\rho_R)a^4}} \;\simeq\; \frac{\alpha^2}{\pi}y^2\lambda^{1/4} \rho_{\rm end}^{3/4} M_P \,,
\eeq
for $R=R_\phi$. In the second equality we have used the fact that, to a good approximation, $\rho_{\phi}a^4 \simeq \rho_{\rm end} \gg \rho_R a^4$ up until the end of reheating. Straightforward integration then yields
\begin{align}
\rho_R \;&\simeq\; \frac{\alpha^2}{\pi}y^2\lambda^{1/4} \rho_{\rm end}^{3/4} M_P \left(\frac{a_{\rm end}}{a}\right)^3\,,\\ \label{eq:Tinstcond}
T\;&=\; \left(\frac{30 \rho_R}{\pi^2 g_{\rm reh}}\right)^{1/4}\,,
\end{align}
that is, $T\propto a^{-3/4}$ during reheating. The end of reheating is determined by the inflaton-radiation equality condition, $\rho_R=\rho_{\phi}$. Solving with the above expression for $\rho_R$, one obtains~\cite{Garcia:2020eof,Garcia:2020wiy}
\begin{align} \label{eq:arehcond}
\frac{a_{\rm reh}}{a_{\rm end}} \;&\simeq\; \frac{\pi \rho_{\rm end}^{1/4}}{\alpha^2y^2 \lambda^{1/4} M_P}\,,\\
T_{\rm reh} \;&\simeq\; \left(\frac{30\lambda}{\pi^6 g_{\rm reh}}\right)^{1/4} \alpha^2 y^2 M_P\,.
\end{align}

\begin{figure}[!t]
\centering
    \includegraphics[width=0.9\textwidth]{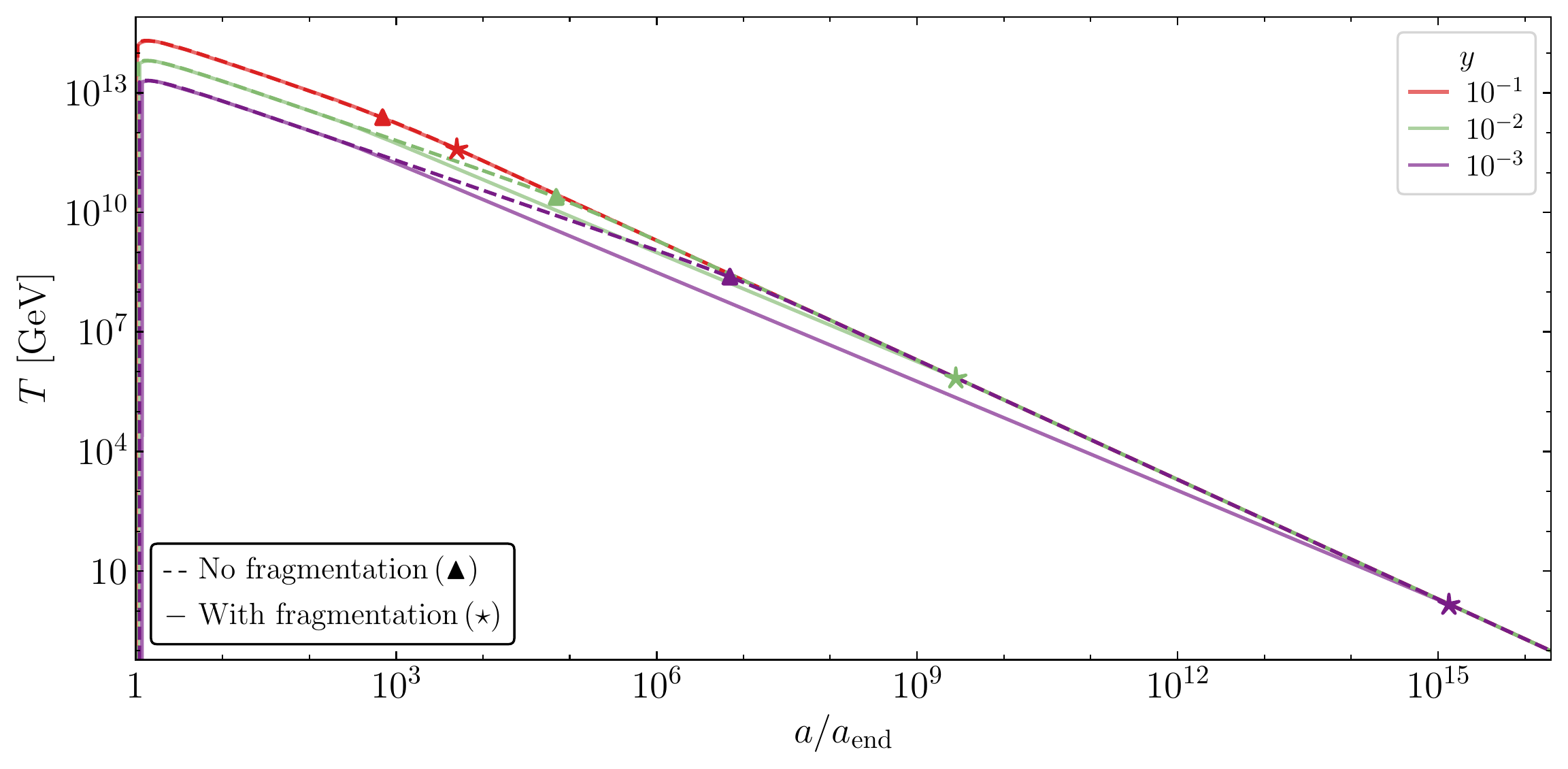}
    \caption{Scale factor dependence of the instantaneous temperature of the relativistic plasma during and after reheating, for a selection of effective inflaton-fermion couplings. Shown as dashed curves is the numerical integration of Eq.~(\ref{eq:continuityradiation}) assuming no fragmentation of the inflaton condensate. For those curves, the end of reheating is marked with a triangle. Continuous curves correspond to the solution of (\ref{eq:continuityradiation}) accounting for the fragmentation of $\phi$. The end of reheating in this case is marked by a star.}
    \label{fig:TvsA}
\end{figure} 

Fig.~\ref{fig:TvsA} shows the evolution of the instantaneous temperature during and after reheating in the condensate approximation, shown as the dashed lines, for $y=\{10^{-1},10^{-2},10^{-3}\}$. This temperature rapidly rises from zero at the end of inflation to reach a maximum $T_{\rm max}$, which may be approximated as~\cite{Garcia:2020wiy}
\beq
T_{\rm max} \;\simeq\;  6\times 10^{14}\, y^{1/2}\,{\rm GeV}\,.
\eeq
Soon after reaching this maximum temperature, the relativistic bath of inflaton decay products continues being populated by these decays, while being redshifted by expansion, following the relation (\ref{eq:Tinstcond}). Finally, upon reaching the end of reheating, given by (\ref{eq:arehcond}), the thermal plasma simply redshifts, $T\propto a^{-1}$ during the radiation dominated epoch.

Fig.~\ref{fig:Treh} shows the reheating temperature $T_{\rm reh}$, as a function of the effective coupling $y$, corresponding to the orange dashed curve in the pure condensate approximation. In the range of couplings selected the reheating temperature is always well above the lower bound imposed by successful big bang nucleosynthesis (BBN), $T_{\rm BBN}\sim 1\,{\rm MeV}$~\cite{Workman:2022ynf}.

\subsubsection{Fragmented inflaton decay}

Taking now into account the preheating phase, we can approximate the total rate $R\simeq R_{\delta \phi}$ at late times, $a/a_{\rm end}\gg \mathcal{O}(10^2)$. Parametrizing
\beq
R_{\delta \phi} \;=\; \frac{y^2}{8\pi}m_{\phi}^2n_{\delta\phi} \;\equiv\;\frac{ \gamma_{\phi}y^2}{8\sqrt{3}\pi}M_P^5 \left(\frac{a_{\rm end}}{a}\right)^{5+x}\,,
\eeq
with $\gamma_{\phi}\simeq 2.49\times 10^{-15}$ and $x\simeq 0.65$ (as determined from the lattice, cf.~Fig.~\ref{fig:rates}), we can immediately integrate (\ref{eq:boltzina}) to obtain
\beq
\rho_R \;\simeq\; \frac{\gamma_{\phi}y^2}{8\pi(1-x)}\left(\frac{M_P^4}{\rho_{\rm end}}\right)^{1/2} M_P^4 \left(\frac{a}{a_{\rm end}}\right)^{-3-x}\,,
\eeq
during reheating. This implies that the instantaneous temperature redshifts as $T\propto a^{-3/4-x/4}\simeq a^{-0.91}$, that is, more rapidly than in the pure-condensate scenario. In this case, reheating ends when
\beq
\frac{a_{\rm reh}}{a_{\rm end}} \;\simeq\; \left[ \frac{8\pi(1-x)}{\gamma_{\phi}y^2}\left(\frac{\rho_{\rm end}}{M_P^4}\right)^{3/2}\right]^{\frac{1}{1-x}}\,,
\eeq
and
\beq
T_{\rm reh} \;\simeq\; \left(\frac{30 \rho_{\rm end}}{\pi^2 g_{\rm reh}}\right)^{1/4} \left[ \frac{\gamma_{\phi}y^2}{8\pi(1-x)}\left(\frac{M_P^4}{\rho_{\rm end}}\right)^{3/2}\right]^{\frac{1}{1-x}}\,,
\eeq
which is the main result of this work. In particular, we note that due to fragmentation, $T_{\rm reh}\propto y^{5.68}$. Therefore, low reheating temperatures are expected even for comparatively large values of $y$ with respect to the pure condensate scenario.

\begin{figure}[!t]
\centering
    \includegraphics[width=0.9\textwidth]{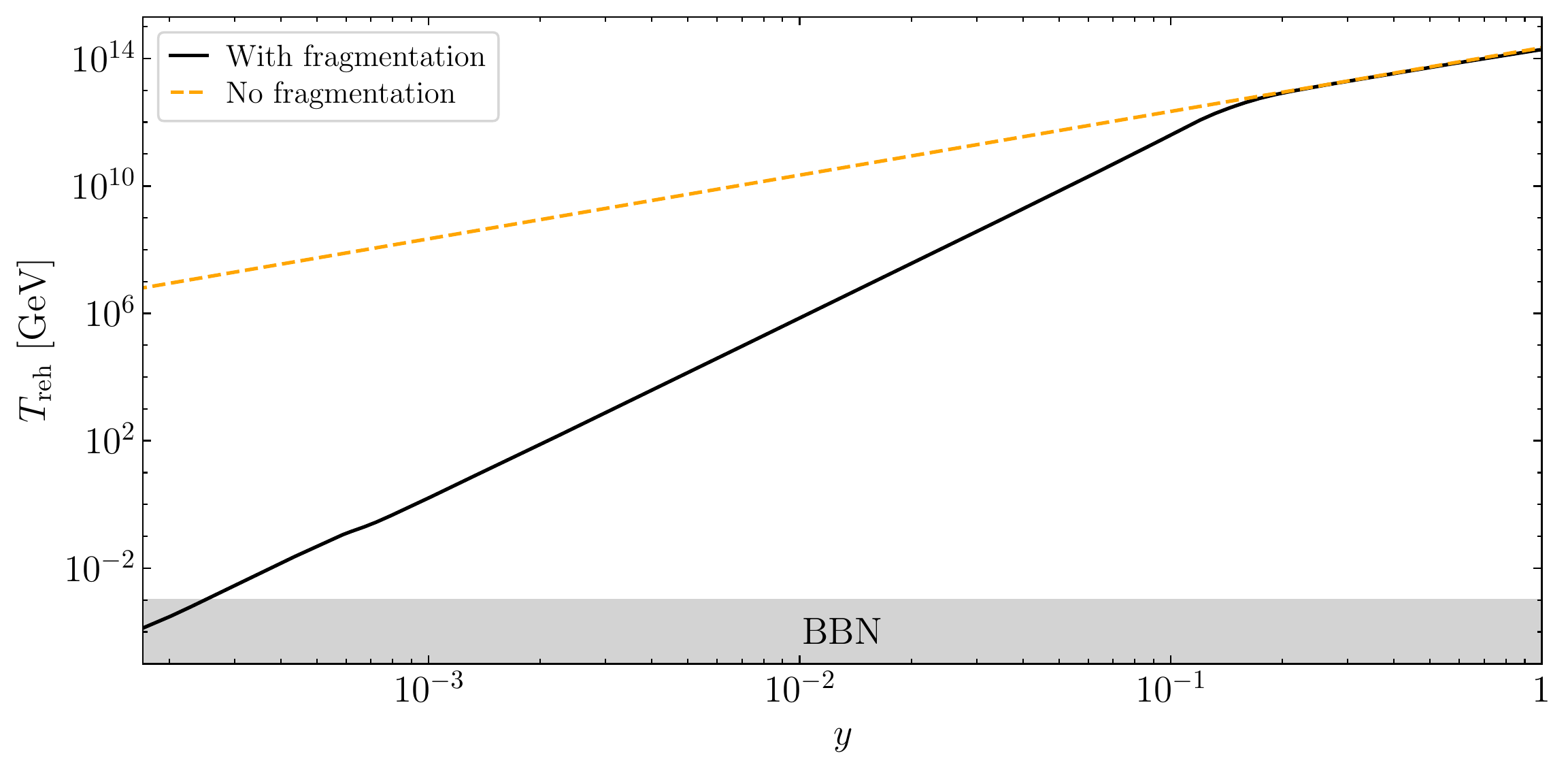}
    \caption{Reheating temperature as a function of the inflaton-fermion coupling, accounting for $\phi$ fragmentation (continuous), and without accounting for fragmentation (dashed). Shown in gray is the region forbidden from big bang nucleosynthesis considerations.}
    \label{fig:Treh}
\end{figure}

The effect of the self-resonance of the inflaton on the evolution of the instantaneous temperature of the relativistic plasma can be clearly seen in Fig.~\ref{fig:TvsA}. Here the continuous curves correspond to the fragmentation scenario. Immediately after the beginning of reheating the temperature follows the condensate result, as the bulk of $\rho_{\phi}$ is still contained in the coherent component of $\phi$. However, after the onset of backreaction, the condensate is depleted and the decay process is dominated by the decay of $k\neq 0$ modes, with a rate that redshifts faster than that of the pure condensate case (cf.~Fig.~\ref{fig:rates}). The result is a faster redshift of $T$, and consequently a reduced reheating temperature, shown in the figure as stars. 

The difference in reheating temperatures can be better appreciated in Fig.~\ref{fig:Treh}. For $y\gtrsim 0.17$ the decay of $\phi$ occurs earlier than the onset of backreaction, and the full result is indistinguishable from the pure condensate scenario. On the other hand, for smaller couplings, $T_{\rm reh}$ is smaller relative to the condensate result, with the difference increasing for decreasing $y$. The small kinks that may be noted in the $T_{\rm reh}$ vs $y$ curve correspond to the effect of the change in the effective number of relativistic degrees of freedom. We finally note that only for $y\gtrsim 2.8\times 10^{-4}$ can the BBN constrain be averted.

\section{Gravitational waves}\label{sec:gws}

In cosmological perturbation theory, tensor metric perturbations are sourced by scalar metric perturbations, starting at second order. This contribution, sensibly negligible prior to fragmentation, would become significant once non-linearity induces sizable mode-mode couplings, efficiently sourcing gravitational waves. In this section we compute the gravitational wave production associated to the large inhomogeneities generated posterior to the end of inflation by the self-fragmentation of the inflaton condensate. Accounting for tensor metric perturbations, the isotropic and homogeneous Friedmann-Robertson-Walker metric is modified to
\begin{equation}
    \diff s^2\,=\,a(\tau)^2\Big[ \diff \tau^2 - \Big( \delta_{ij} + h_{ij} \Big) \diff x^i \diff x^j   \Big]\,,
\end{equation}
where the transverseness and tracelessness (TT) conditions $\partial_i h_{ij}=0$ and $h^i_i=0$ are satisfied. The transverse-traceless component of the anisotropic stress $\Pi^\text{TT}_{ ij } = \big[\partial_i \phi \partial_j \phi \big]^\text{TT}$ appears as a source term in the equation of motion for the tensor metric perturbations\footnote{$h_{ij}$ represent two independent tensor degrees of freedom satisfying the same equation.}
\begin{equation}
h^{\prime \prime}_{ij}(\bk,\tau)+ 2 \mathcal{H} h^{\prime }_{ij}(\bk,\tau)+k^2 h_{ij}(\bk,\tau) \, = \, \dfrac{2}{M_P^2} \Pi^\text{TT}_{ ij } (\bk,\tau)\,,
\end{equation}	
where the anisotropic stress projected onto the transverse-traceless component in Fourier space is~\cite{Dufaux:2007pt}
\begin{equation}
\Pi^\text{TT}_{ ij } (\bk,\tau) \, = \left( \, P_{i\ell}(\hat \bk) P_{jm}(\hat \bk)-\dfrac{1}{2} P_{ij}(\hat \bk) P_{\ell m}(\hat \bk) \right) \Pi_{ \ell m } (\bk,\tau)\,,
\end{equation}
with $P_{ij}=\delta_{ij}-\hat{k_i}\hat{k_j}$ and $\hat k_i=k_i/k$ and
\begin{equation}
     \Pi_{ \ell m } (\bk,\tau) \, = \, \int\dfrac{\diff^3 \bp }{(2 \pi)^{3/2}} \, p_\ell \, p_m \, \phi(\bp,\tau) \, \phi(\bk-\bp,\tau) \,.
\end{equation}\par \medskip

\subsection{Simulated spectrum}

\begin{figure}[!t]
\centering
    \includegraphics[width=0.47\textwidth]{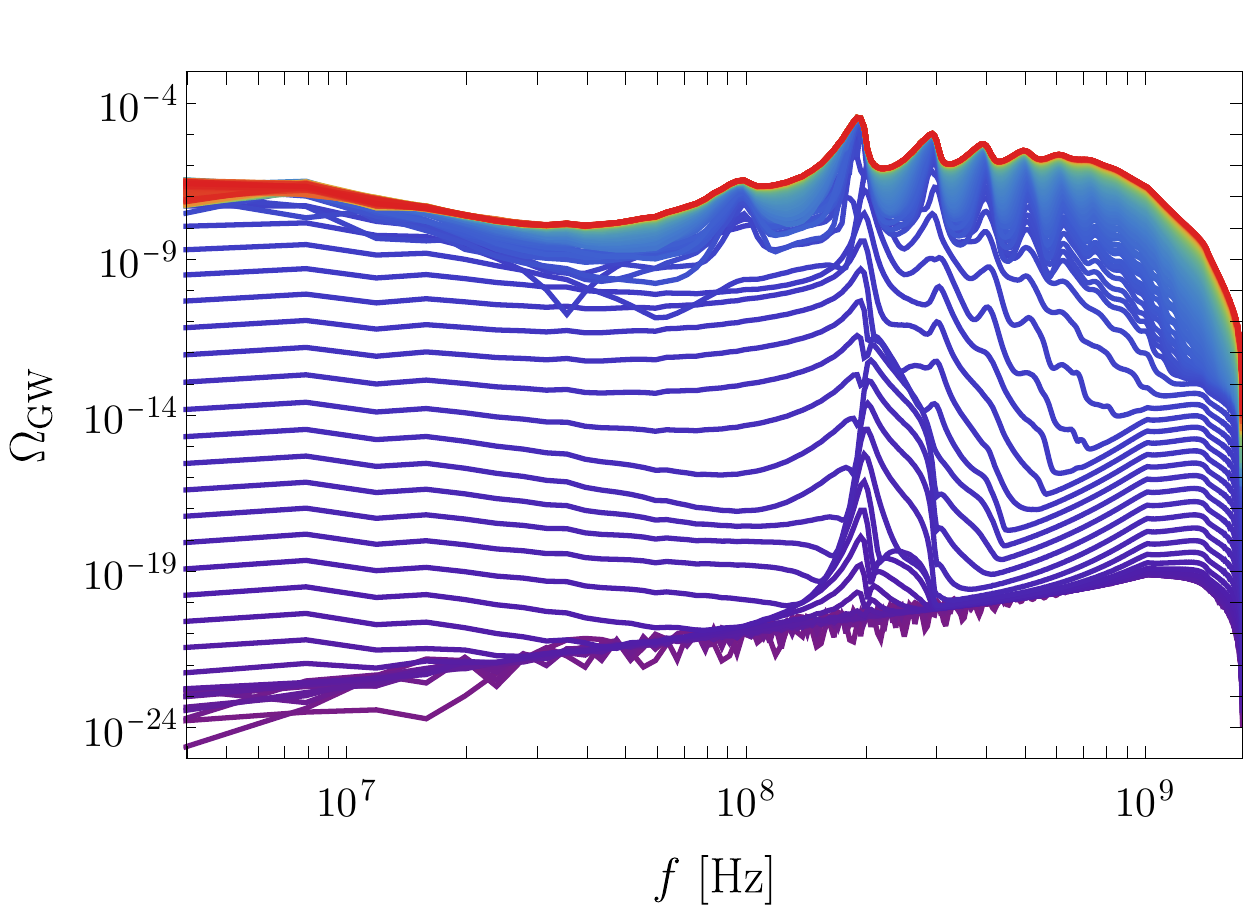}\hfill
       \includegraphics[width=0.476\textwidth]{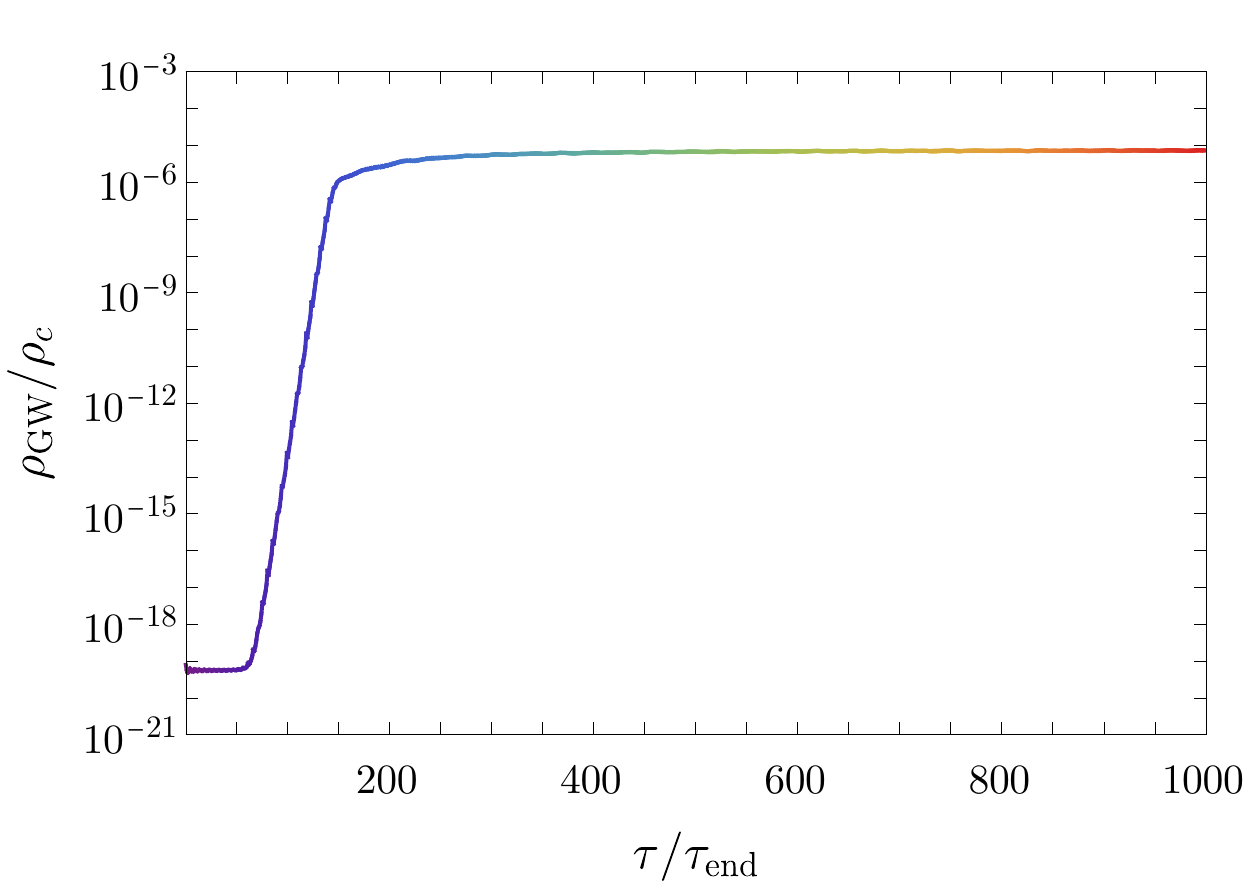} 
    \caption{\textbf{Left:} Differential gravitational wave energy density per logarithmic (present day)  frequency interval, normalized to the critical density, evaluated at selected conformal time. The color code corresponds to the evaluation time of the right panel. \textbf{Right:} Integrated differential GW energy density evolution with respect to conformal time, normalized to the critical energy density.}
    \label{fig:GW}
\end{figure}

We use \CL~to simulate the production of gravitational waves for our model. In practice, a TT projector is implemented in a discretized version on the lattice. The non-uniqueness of the definition of such a projector can induce small differences in the GW spectrum. However, it has been argued it only marginally affects the UV part of the GW spectrum~\cite{Figueroa:2011ye,Figueroa:2020rrl}. On the lattice, the GW energy density is computed as a volume average
\begin{equation}
\rho_\text{GW} \, = \, \dfrac{M_P^2}{4 a^4}\dfrac{1}{V} \int \diff^3 \bk \, \bar{h}_{ij}^\prime (\tau,\bk) \, \bar{h}_{ij}^{\prime *}(\tau,\bk) \,,
\end{equation}
with $\bar{h}_{ij}=a h_{ij}$. This sum is performed by \CL, after discretization, over a finite lattice volume $V$ (details can be found in Ref.~\cite{Figueroa:2021yhd}). The normalized GW energy density per logarithmic frequency interval can be expressed as
\begin{equation}
\Omega_\text{GW} (k) \, = \, \dfrac{1}{\rho_c} \dfrac{\diff  \rho_\text{GW}}{\diff \log k}\,,
\end{equation}
where $\rho_\text{GW}$ is the total GW energy density and $\rho_c$ is the (time-dependent) critical energy density defined via $\rho_c = 3 H^2M_P^2 $. The GW frequency $f$ at the present epoch can be related to the comoving Fourier scale $k$ via~\cite{Dufaux:2007pt}
\begin{equation}
f \, = \, \left( \dfrac{a_\text{end}}{a_0} \right) \dfrac{k}{2 \pi} \, = \, \left( \dfrac{\rho_{\text{rad},0}}{\rho_\text{end}} \right)^{1/4} \left( \dfrac{g_\text{reh}}{g_0} \right)^{1/4} \left( \dfrac{g_{s,0}}{g_{s,\text{reh}}} \right)^{1/3}  \dfrac{k}{2 \pi} \, \simeq \, 1.46 \times 10^8 \, \left( \dfrac{k}{0.7 \,m_\text{end}} \right) ~\text{Hz} \,,
\label{eq:translatektof}
\end{equation}
where we normalized the scale $k$ to the resonant value from the Floquet analysis $k \simeq 0.7 \,m_\text{end}$.  $\rho_{\text{rad},0}$ is the radiation energy density at the present time. The GW energy density estimated from our lattice simulation is shown in Fig.~\ref{fig:GW}, evaluated a different times on the left panel, as a function of the frequency, and integrated over frequency as a function of time on the right panel. From this figure one can see that the GW energy density significantly increases at around $\tau/\tau_\text{end}\simeq 60-80$ when scalar inhomogeneities, triggered by the parametric resonance discussed in Sec.~\ref{sec:parametricresonance}, start to increase, as seen in Fig.~\ref{fig:rhos}. The GW energy density relative to the critical energy density stabilises at around $\tau/\tau_\text{end}\simeq 175$ when fragmentation is achieved, asymptoting smoothly towards a constant value $\rho_\text{GW}/\rho_c \, \simeq \, 7.3 \times 10^{-6}$ at larger $\tau/\tau_\text{end}$.

\begin{figure}[!t]
\centering
    \includegraphics[width=\textwidth]{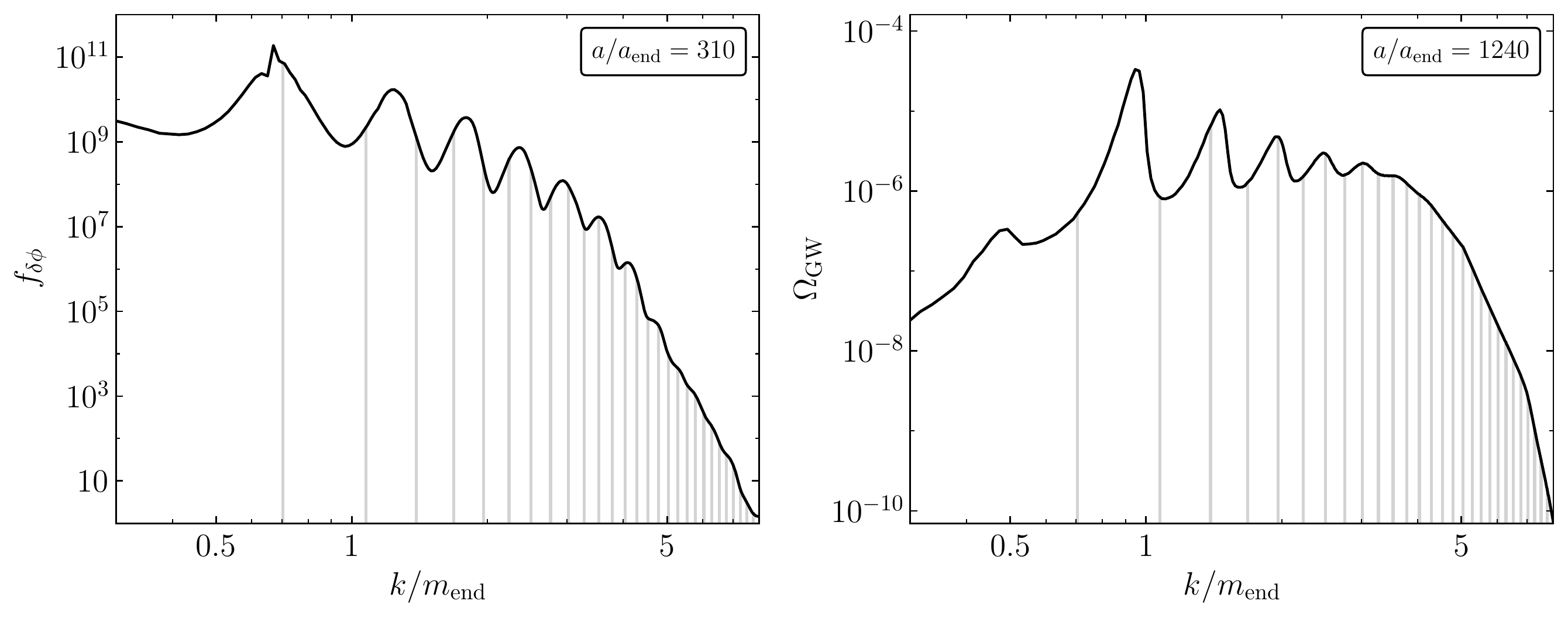}
    \caption{Comparison between the peak structure of the scalar phase space distribution at the onset of fragmentation (left), and of the final gravitational wave spectrum (right), with the predictions of the Boltzmann approach (vertical gray lines). The Boltzmann lines are only indicative of the value of the comoving momenta, the ordinate values are matched with the corresponding black curve.}
    \label{fig:boltz}
\end{figure}

One of the most striking features of the final GW spectrum, i.e.~the red curve in Fig.~\ref{fig:GW} on the left panel, is the peak structure.\,\footnote{Such structure was already identified for a quartic potential in~\cite{Fu:2017ero}.} In the red curve, the dominant peak is located at a frequency $f\simeq1.9\times 10^8~\text{Hz}$ slightly larger than the predicted resonant comoving scale $k/m_\text{end} \simeq 0.7$ from the Floquet analysis from Eq.~(\ref{eq:translatektof}). From the left panel of Fig.~\ref{fig:GW}, one can see that the initial GW spectra dominant peak was located at a slightly lower frequency and progressively displaced towards higher frequencies, explaining such difference. Disregarding the smaller peak around $\sim 10^8~ \text{Hz}$ that seems to appear at the onset of fragmentation (originated by non-linear effects), we can compare the peak structure of the spectrum to those of the scalar spectra and predictions from the Boltzmann approach, cf. Eq.~(\ref{eq:Boltzmannpeakprediction}). A selection of numerical results are presented in Table~\ref{tab:my_table1}, and are shown graphically in Fig.~\ref{fig:boltz} for the inflaton fluctuation PSD and the gravitational wave spectrum. Even if the peak structure of the scalar spectra are smoothed out by the end of reheating, the structure of the peaks remains imprinted in the tensor spectra. The Boltzmann approach allows for a very decent prediction for the location of the peaks in the scalar spectra. More remarkably, the Boltzmann approach allows to explain the spacing between the first peaks with a precision at the $<10\%$ level. \par \medskip

\begin{table}[h!] \vspace{0.3cm}
    \centering
    \begin{tabular}{c|c||c|c|c|c|c}
   & $\hat{q}_1$  & $\hat{q}_1/\hat{q}_2$  &  $\hat{q}_1/\hat{q}_3$   &   $\hat{q}_1/\hat{q}_4$ &$\hat{q}_2/\hat{q}_4$ &$\hat{q}_2/\hat{q}_5$ 
\\
        \hline
        \hline
  Lattice  (scalar)   & 0.7  & 0.54 &  0.48 & 0.3 & 0.56 & 0.45  \\ \hline
    Lattice  (tensor)   & 0.91   & 0.65 & 0.48  & 0.38 & 0.59 & 0.48  
  \\ \hline
       Boltzmann & 0.7   &  0.65  & 0.51  & 0.42 & 0.63 & 0.54 \\  \hline \hline
      Linear/Hartree  &  0.7  & - & - & - & - & -\\  \hline
      Floquet  &  0.7  & - & - & - & - & -\\  \hline
    \end{tabular}
    \caption{Comparison of the peak structure of the scalar and tensor spectra where $\hat{q}_i$ corresponds to the location of the $i^\text{th}$ peak in terms of the rescaled momentum $q\equiv k /m_\text{end}$. The Boltzmann predictions are based on Eq.~(\ref{eq:Boltzmannpeak}) and Eq.~(\ref{eq:Boltzmannpeakprediction}).}
    \label{tab:my_table1}
\end{table}

\subsection{Signal and detection prospects}

There are essentially two distinct GW signals corresponding to two different physical processes predicted by this model. First, during inflation quantum fluctuations of tensor modes are stretched on macroscopic super-horizon scales. Once such scales re-enter the horizon in a subsequent phase of the universe, tensor perturbations manifest as a stochastic background of gravitational waves. Primordial tensor perturbations, customarily parametrized in terms of the tensor-to-scalar ratio can be expressed 
for the potential of Eq.~(\ref{eq:phipotential}) as~\cite{Garcia:2020wiy}
\begin{equation}
    r \, \simeq \, \dfrac{12}{N_*^2} \, \simeq \, 3.8 \times 10^{-3} \, \left( \dfrac{56}{N_*} \right)^2 \,,
\end{equation}
evaluated at the CMB fiducial scale $k_*=0.05~\text{Mpc}^{-1}$. Such value for the tensor-to-scalar ratio appears to be close to the sensitivity reach for the upcoming Simons Observatory (SO) $r<6 \times 10^{-3}$~\cite{SimonsObservatory:2018koc}. Moreover, future missions such as LiteBIRD and CMB Stage-4 (CMB-S4) should reach sensitivities as small as $r<2\times 10^{-3}$~\cite{LiteBIRD:2022cnt}  and $r< 10^{-3}$~\cite{SO-CMBS4} respectively, allowing to disprove or confirm this model.

\begin{figure}[!t]
\centering
    \includegraphics[width=0.7\textwidth]{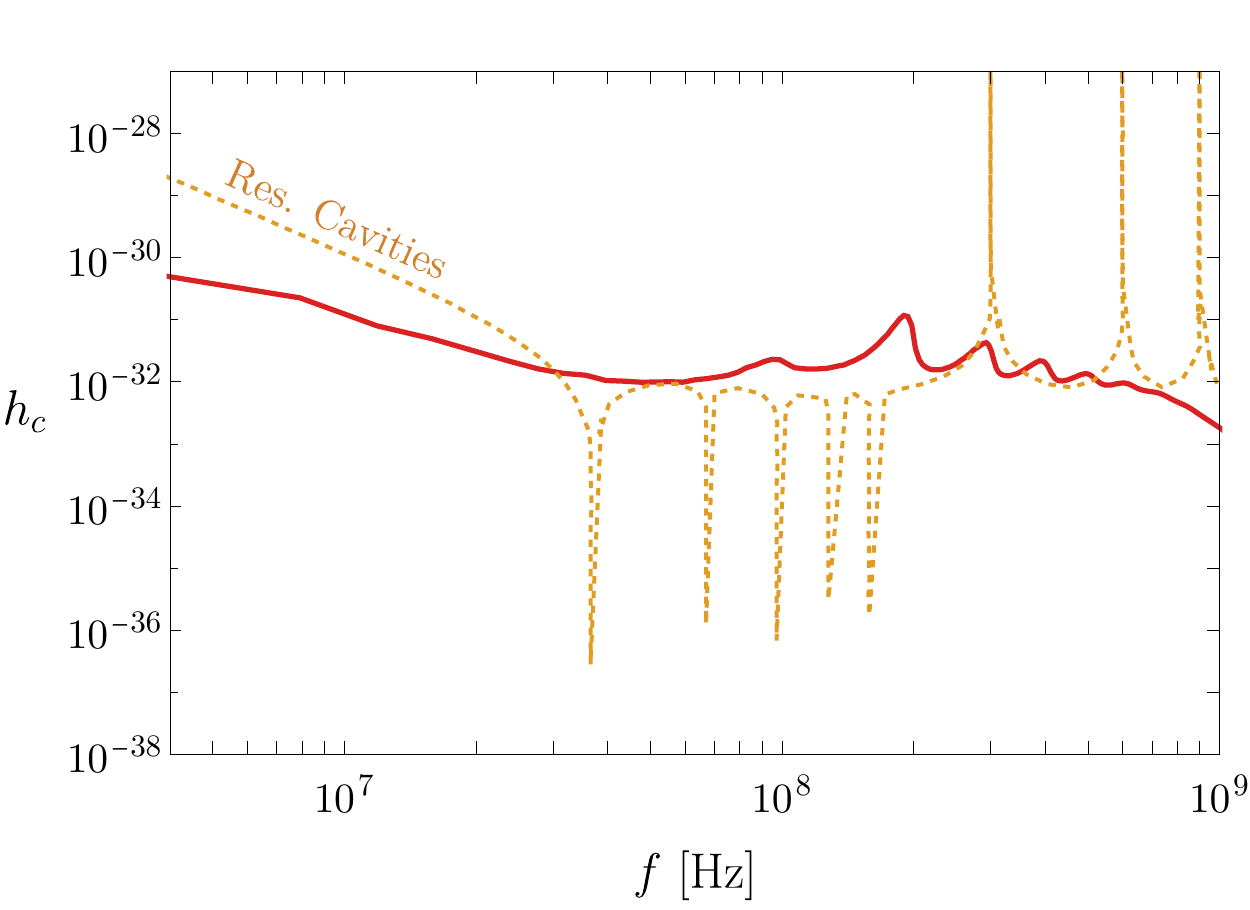}
    \caption{Simulated amplitude of the present-day GW signal amplitude as a function of the frequency (solid red line) and sensitivity estimate for resonant electromagnetic cavities of Ref.~\cite{Herman:2022fau} (orange dashed line).}
    \label{fig:GWstrain}
\end{figure}

The second GW signal is generated during fragmentation from large inflaton inhomogeneities which we simulated using \CL. We can estimate the current GW energy density from
\begin{equation}
\rho_\text{GW,0} \, = \, \rho_\text{GW,reh} \left( \dfrac{a_\text{reh}}{a_0}\right)^4 \, = \,  \rho_\text{GW,reh}  \left( \dfrac{\rho_{\text{rad},0}}{\rho_{c,\text{reh}}} \right) \left( \dfrac{g_\text{reh}}{g_0} \right) \left( \dfrac{g_{s,0}}{g_{s,\text{reh}}} \right)^{4/3}\,,
\end{equation}
where we took $g_\text{reh}=g_{s,\text{reh}}=106.75$, $g_{s,0}=3.909$ and $g_0=3.363$. The present day normalized GW energy density can thus be expressed as
\begin{equation}
\Omega_{\text{GW},0} (f) \, = \, \dfrac{\rho_{\text{rad},0}}{\rho_{c,0}}  \left( \dfrac{g_\text{reh}}{g_0} \right) \left( \dfrac{g_{s,0}}{g_{s,\text{reh}}} \right)^{4/3}   \, \Omega_\text{GW,reh}(f) \, \simeq \,   \, 2 \times 10^{-5} \, \Omega_\text{GW,reh}(f) \,.
\end{equation}
Such gravitational wave signal would induce a deviation to the effective number of relativistic species $\Delta N_\text{eff}\equiv N_\text{eff}-N_\text{eff}^\text{SM}$ with $N_\text{eff}^\text{SM}=3.046$. This contribution can be expressed as
\begin{equation}
\Delta N_\text{eff} \, = \, \dfrac{8}{7} \left(\dfrac{11}{4} \right)^{4/3}  \left(\dfrac{\rho_\text{GW,0}}{\rho_{\text{rad},0}} \right) = \, \dfrac{8}{7} \left(\dfrac{11}{4} \right)^{4/3}  \left( \dfrac{g_\text{reh}}{g_0} \right) \left( \dfrac{g_{s,0}}{g_{s,\text{reh}}} \right)^{4/3} \left(\dfrac{\rho_\text{GW,reh}}{\rho_{c,\text{reh}}} \right) \, \simeq \, 1.7  \left(\dfrac{\rho_\text{GW,reh}}{\rho_{c,\text{reh}}} \right)\,.
\end{equation}
We find that for $\tau/\tau_\text{end}>300$, as illustrated on the right panel of Fig.~\ref{fig:GW}, the produced GW energy density asymptotes during reheating to the value $\rho_\text{GW}/\rho_c \, \simeq \, 7.3 \times 10^{-6}$. This allows to estimate the contribution to the effective number of non-relativistic degrees of freedom
\begin{equation}
\Delta N_\text{eff} \,  \simeq \, 1.2 \times 10^{-5}  \,,
\end{equation}
which is far below the current sensitivity from Planck $N_\text{eff}=2.98^{+0.39}_{-0.38}$ at $95\%$ confidence level (TT, TE, EE+lowE+lensing+BAO)~\cite{Planck:2018vyg} as well as future CMB missions CMB-S4~\cite{Abazajian:2019eic} and CMB-HD~\cite{CMB-HD:2022bsz} with an expected precision for $\Delta N_\text{eff}$ of $0.06$ and $0.027$ respectively at $95\%$ confidence level. In order to facilitate comparison with sensitivity prospects for future experiments, we can define the dimensionless characteristic strain as
\begin{equation}
h_c(f) \, = \, \sqrt{\dfrac{3 H_0^2 \Omega_{\text{GW,0}}(f)}{2 \pi^2}} \dfrac{1}{f} \,.
\end{equation}
In Fig.~\ref{fig:GWstrain} we compare our estimate of the present day strain as a function of the frequency, in comparison to the sensitivity estimate for high-frequency resonant electromagnetic cavities of Ref.~\cite{Herman:2022fau}. The frequency range of the GW signal emitted from the fragmentation process falls in the appropriate sensitivity range of the proposal of Ref.~\cite{Herman:2022fau}. Remarkably, the reconstitution of the peak structure of such high-frequency GW signal, in addition to measurements of the tensor-to-scalar ratio would help break the ambiguities in determining the realization of inflation.

\section{Summary and conclusions}
\label{sec:conclusions}

In this work we have investigated the post-inflationary dynamics of the inflaton field and its decay products, under the assumption of a potential with a quartic minimum. For definiteness, we considered the T-model of inflation, compatible with current constraints on inflation. After inflation, the inflaton condensate coherently oscillates about its minimum. Triggered by parametric resonance effects sourced by the oscillating background, exponential growth of inflaton inhomogeneities result in the fragmentation of this condensate. In this work we explored the role and consequences of such non-linear effects on the post-inflationary history and the successful reheating of the universe. We summarize in the following the key aspects and results of our analysis. \par \medskip

\noindent
\textbf{Parametric resonances and fragmentation.} First, we characterized the parametric resonant effects triggering fragmentation using a standard Floquet approach. We found that due to the conformal symmetry of the system, only Fourier scales located at $k/m_\text{end}\simeq 0.7$ would experience a significant exponential growth. By solving numerically the equation of motion for the inflaton fluctuations at linear order in perturbation theory, we were able to observe a growth of modes with $k/m_\text{end}\simeq 0.7$, therefore confirming predictions from the Floquet analysis. To go beyond the linear analysis, we simulated the evolution of the space-time dependent inflaton field configuration with the nonperturbative code~\CL. By computing the occupation number for the inflaton perturbations, we recovered results from the linear approach in the first instants after the end of inflation. We find that at around 5 $e$-folds after the end of inflation, perturbations enter the non-linear regime, efficiently backreacting on the inflation condensate, leading to fragmentation. 

Before rescatterings fully redistribute the energy density into other modes, the occupation number of inflaton quanta features several peaks located at higher momentum than the resonant $k/m_\text{end}\simeq 0.7$ Floquet mode. We estimated analytically the position of these peaks by means of a Boltzmann approach. Notably, despite the validity of this formalism only in the linear regime, we find that it not only accurately predicts the dominant Floquet peak, but it also matches the lattice peak location within a few percent, at least before they are fully washed-out by rescatterings.

Importantly, we found that the redistribution of the inflaton energy density does not fully erase the coherent zero mode. Upon the onset of fragmentation, the energy budget of the universe becomes dominated by a collection of inflaton quanta redshifting as radiation $\rho_{\delta \phi} \propto a^{-4}$, but a leftover inflaton condensate persists with $\overbar{\rho_\phi} \propto a^{-5.3}$. 
\par \medskip

\noindent
\textbf{Reheating.} The main result of this work consists in the exploration of the consequences of fragmentation to successfully achieve reheating; that is, the transition to a universe dominated by radiation in thermal equilibrium. By using a Boltzmann approach, we computed the contributions to the production rates of relativistic fermionic states coming from the oscillating inflaton condensate $R_\phi$, and also from the fragmentated population of inflaton quanta $R_{\delta \phi}$. Prior to backreaction, the rate at which energy is injected into the thermal bath per unit volume per unit time, scales as $R_{\phi}\propto a^{-5}$. After fragmentation, we find that the condensate rate $R_{\phi} \propto a^{-6.6}$, and the inflaton particle rate $R_{\delta \phi} \propto a^{-5.65}$, both redshift faster, due to the rapid decrease in the time-dependent induced inflaton effective mass. As a consequence, reheating is less efficient after fragmentation. The instantaneous temperature falls as $T\propto a^{-0.91}$ (compared to $T\propto a^{-3/4}$ without backreaction). For fermionic decays, the reheating temperature scales as $T_{\rm reh}\propto y^{2}$ for $y\gtrsim 10^{-1}$, and as $T_{\rm reh}\propto y^{5.68}$ for $y\lesssim 10^{-1}$, where $y$ is the inflaton-matter Yukawa coupling. Requiring the reheating temperature to be larger than the BBN temperature $T_\text{BBN} \sim\,\text{MeV}$ necessitates couplings larger than $y>2.8 \times 10^{-4}$.

We must emphasize that these results are valid only for a quartic inflaton minimum. Nevertheless, the lattice+Boltzmann formalism developed in this work is general, and can be applied to a more diverse collection of inflation models, with non-quartic potentials. The precise details of the evolution of energy densities and temperatures will vary on a model-by-model basis, and will be the subject of future follow-up work.

\par \medskip

\noindent
\textbf{Tensor perturbations.} The large mode-mode couplings for the inflaton inhomogeneities inducing fragmentation can also source sizable tensor perturbations. We estimated the production of gravitational waves from a simulation with~\CL. The gravitational wave spectrum features several peaks located at high frequencies $f \sim 10^{8}-10^{9}~\text{Hz}$ whose precise locations are inherited from the scalar spectrum. The ratios of the location for the various peaks matches relatively well predictions from the Boltzmann approach. We find that the GW energy-density contribution to the effective number of relativistic species $\Delta N_\text{eff} \sim 10^{-5}$ is beyond the reach of any upcoming experiments. However, the frequency range for the GW spectrum appears in sensitivity-reach estimate for resonant electromagnetic cavities~\cite{Herman:2022fau}.

\par \medskip

\begin{acknowledgments}

We would like to thank Keith Olive, Yann Mambrini, Andreas Ringwald and Sarunas Verner for helpful discussions. MG is supported by the DGAPA-PAPIIT grant IA103123 at UNAM, and the CONAHCYT ``Ciencia de Frontera'' grant CF-2023-I-17. MP acknowledges support by the Deutsche Forschungsgemeinschaft (DFG, German Research Foundation) under Germany's Excellence Strategy – EXC 2121 “Quantum Universe” – 390833306. This work was made possible by the support of the Institut Pascal at Universite Paris-Saclay during the Paris-Saclay Astroparticle Symposium 2022, with the support of the P2IO Laboratory of Excellence (program “Investissements d’avenir” ANR-11-IDEX-0003-01 Paris-Saclay and ANR-10-LABX-0038), the P2I axis of the Graduate School Physics of Universite Paris-Saclay, as well as IJCLab, CEA, IPhT, APPEC, the IN2P3 master projet UCMN and EuCAPT ANR-11-IDEX-0003-01 Paris-Saclay and ANR-10-LABX-0038). Non-perturbative numerical results in the Hartree approximation were obtained from a custom Fortran code utilizing the thread-safe arbitrary precision package MPFUN-For~\cite{mpfun}.
\end{acknowledgments}

\addcontentsline{toc}{section}{References}
\bibliographystyle{utphys}
\bibliography{references}

\end{document}